\documentclass[nolinenum]{jfp}

\usepackage[T1]{fontenc}
\usepackage{ascii}

\usepackage{stmaryrd}

\usepackage[frozencache]{minted}
\setminted{fontsize=\small}
\definecolor{mygray}{rgb}{0.95,0.95,0.95}

\usepackage{tikz}

\usepackage{extarrows}
\usepackage{xhfill}

\DeclareMathOperator{\supp}{supp}

\definecolor{purple}{rgb}{1.0,0,1.0}

\def\us{\char`\_}

\def\coqin#1{\text{\mintinline{ssr}{#1}}}
\def\coqgen{\textsc{CoqGen}}
\def\monae{\textsc{Monae}}
\def\coq{\textsc{Coq}}
\def\ocaml{OCaml}
\def\ssreflect{\textsc{SSReflect}}
\def\mathcomp{\textsc{MathComp}}
\def\hb{\textsc{Hierarchy-Builder}}
\def\newterm#1{\textsl{#1}}
\def\isdef{\overset{\mathit{def}}{=}}

\begin{document}

\journaltitle{Manuscript}
\cpr{no copyright information here}
\doival{see arxiv}

\lefttitle{manuscript}
\righttitle{manuscript}

\totalpg{\pageref{lastpage01}}
\jnlDoiYr{2023}

\title{A Practical Formalization of Monadic Equational Reasoning in
  Dependent-type Theory\footnote{This manuscript is based on the revision
  of two papers presented at the MPC conference
  \citep{affeldt2019mpc,saito2022mpc} and on work presented at the Coq Workshop \citep{saikawa2023coq}.
}
}

\begin{authgrp}
\author{Reynald  Affeldt}
\affiliation{National Institute of Advanced Industrial Science and Technology (AIST),\\
        Digital Architecture Research Center, Tokyo, Japan\\
        (\email{reynald.affeldt@aist.go.jp})}

\author{Jacques Garrigue}
\affiliation{Nagoya University,\\
        Graduate School of Mathematics, Japan\\
        (\email{garrigue@math.nagoya-u.ac.jp})}

\author{Takafumi Saikawa}
\affiliation{Nagoya University,\\
        Graduate School of Mathematics, Japan\\
        (\email{tscompor@gmail.com})}
\end{authgrp}


\begin{abstract}
One can perform equational reasoning about computational effects with
a purely functional programming language thanks to monads. Even though
equational reasoning for effectful programs is desirable, it is not
yet mainstream. This is partly because it is difficult to maintain
pencil-and-paper proofs of large examples.
We propose a formalization of a hierarchy of effects using monads in
the Coq proof assistant that makes monadic equational reasoning
practical.
Our main idea is to formalize the hierarchy of effects and algebraic
laws as interfaces like it is done when formalizing hierarchy of
algebras in dependent type theory.
Thanks to this approach, we clearly separate equational laws from
models. We can then take advantage of the sophisticated rewriting
capabilities of Coq and build libraries of lemmas to achieve concise
proofs of programs. We can also use the resulting framework to
leverage on Coq's mathematical theories and formalize models of
monads.
In this article, we explain how we formalize a rich hierarchy of
effects (nondeterminism, state, probability, etc.), how we mechanize
examples of monadic equational reasoning from the literature, and how
we apply our framework to the design of equational laws for a subset
of ML with references.
\end{abstract}

\maketitle

\section{Introduction}
\label{sec:introduction}

Pure functional programs are suitable for equational reasoning because
they are referentially transparent. Monadic equational reasoning
\citep{gibbons2011icfp} is an approach to reason equationally about
programs with side effects using the algebraic properties of
monads. In this approach, effects (such as state, nondeterminism,
probability, etc.) are defined by interfaces with a set of equations
and these interfaces can be combined and extended to represent the
combination of several effects. A number of programs using combined
effects have been verified in this way
\citep{gibbons2011icfp,oliveira2012jfp,chen2017netys,shan2018tutorial,mu2019tr2,pauwels2019mpc}
and this approach is also applicable to program derivation
\citep{mu2019tr3,mu2020flops}.  Among these experiments, some have
been formally verified with proof assistants based on dependent type
theory such as \coq{} and Agda.

The implementation of monadic equational reasoning in proof assistants
based on dependent type theory raises a number of issues.
The main issue is the construction of the hierarchy of monad
interfaces.  There exist several approaches such as the type classes
and canonical structures~\citep[Chapter Canonical
Structures]{coq}. The construction of a hierarchy of interfaces
nevertheless requires care because large hierarchies are known to
suffer scalability issues due to the complexity of type inference
\citep[Sect.~2.3]{garillot2009tphols}.
In the context of monadic equational reasoning, shallow embedding is
the privileged way to represent monadic functions. This is the source
of other issues. For example, when needed, induction w.r.t.\ syntax
requires the use of reflection. The use of shallow embedding also
makes it important to know the techniques to deal with non-structural
recursion. These are general concerns in proof assistants but in the
context of monadic equational reasoning they can prevent stating a
proof goal or key reasoning steps.

In this paper, we explain the implementation and the practice of a
\coq{} library called \monae{} that provides support for formal
verification based on monadic equational reasoning.
We leverage on existing tools for the \coq{} proof assistant:
\hb~\citep{cohen2020fscd}, a tool intended at the formalization of
mathematical structures that we use to formalize an extensible
hierarchy of interfaces, and \ssreflect{} \citep{gonthier2010jfr}
\citep[Chapter The SSReflect proof language]{coq}, an extension of
\coq{} that provides more versatile rewriting tactics.
\monae{} has already proved useful by uncovering errors in
pencil-and-paper proofs (e.g., \citep[Sect. 4.4]{affeldt2019mpc}),
leading to new fixes for known errors (e.g.,
\citep{affeldt2020types}), and providing clarifications for the
construction of monads used in probabilistic programs (e.g.,
\citep[Sect.\ 6.3.1]{affeldt2021jfp}).
In this article, our goal is to summarize the technical aspects of
\monae{}: how we formalize interfaces of effects and model of monads,
how we set up the formalization of an existing program derivation, and
how we investigate the design of a new interface for an ML-like
language with references.

\paragraph*{Illustrating example: fast product}
We explain monadic equational reasoning in \monae{} using an example
by \cite{gibbons2011icfp} that shows the equivalence between a
functional implementation of the product of integers (namely,
\coqin{product}) and a monadic version (\coqin{fastprod}).
On the left of Fig.~\ref{fig:fastprod}, we (faithfully) reproduce the
series of rewritings that constitute the original proof
\citep[Sect.~5.1]{gibbons2011icfp}.  On the right, we display the
equivalent series of \coq{} goals and tactics in \monae.

In \coq{}, the \coqin{product} of natural numbers is simply defined as
\coqin{foldr muln 1}.
The ``faster'' product \coqin{fastprod} can be implemented using the failure monad and
the exception monad as follows:
\begin{minted}[bgcolor=mygray,numbers=left,xleftmargin=1.5em,escapeinside=88]{ssr}
Definition work (M : failMonad) s : M nat :=
  if O \in s then fail else Ret (product s). 8\label{line:Ret}8
Definition fastprod (M : exceptMonad) s : M nat :=
  catch (work s) (Ret O).
\end{minted}
The notation \coqin{Ret} (line~\ref{line:Ret}) corresponds to the unit of the monad.  The
function \coqin{work} uses the \coqin{fail} operator of the failure
monad \coqin{failMonad} and \coqin{fastprod} uses the \coqin{catch}
operator of the exception monad \coqin{exceptMonad}.
We observe that the user can write a monadic program using one monad
(\coqin{failMonad} or \coqin{exceptMonad}) and still use a notation to
refer to the unit operator (\coqin{Ret}) of the base monad,
illustrating subtyping.
The formalization of the relevant monads and their algebraic laws is
explained in Sections \ref{sec:formalization_monads},
\ref{sec:nondeterminism}, and~\ref{sec:exceptMonad}, but the details
are not important to understand the illustrating example.

\def\inbra#1{=$\llbracket$ #1 $\rrbracket$}
\def\hask#1{\mintinline[fontsize=\small]{haskell}{#1}}

\begin{figure}
\begin{tabular}{l|l}
Pencil-and-paper proof & \coq{} intermediate goals and tactics \\
\cite[\S 5.1]{gibbons2011icfp} & using \monae{} \\
\hline
\hask{fastprod xs}                    & \coqin{fastprod s} \\
\inbra{definition of \hask{fastprod}} & \inbra{\coqin{rewrite /fastprod}} \\
\hask{catch (work xs) (ret 0)}     & \coqin{catch (work s) (Ret 0)} \\
\inbra{specification of \hask{work}}  & \inbra{\coqin{rewrite /work}}\\
\hask{catch (if 0 in xs then fail}    & \coqin{catch (if 0 \in s then fail} \\
\hask{else ret (product xs)) (ret 0)} & \coqin{else Ret (product s)) (Ret 0)} \\
\inbra{lift out the conditional}      & \inbra{\coqin{rewrite lift_if if_ext}} \\
\hask{if 0 in xs then catch fail (ret 0)} & \coqin{if 0 \in s then catch fail (Ret 0)} \\
\hask{else catch (ret (product xs)) (ret 0)} & \coqin{else catch (Ret (product s)) (Ret 0)} \\
\inbra{laws of catch, fail, and ret} & \inbra{\coqin{rewrite catchfailm catchret}}\\
\hask{if 0 in xs then ret 0} & \coqin{if 0 \in s then Ret 0} \\
\hask{else ret (product xs) } & \coqin{else Ret (product s)} \\
\inbra{arithmetic: 0 in xs $\Rightarrow$ \hask{product xs} = 0} & \inbra{\coqin{case: ifPn => // /product0}}\\
\hask{if 0 in xs then ret (product xs)} & (\coqin{product0}$\isdef\forall s.\; 0 \in s \to \coqin{product}\, s = 0$) \\
\hask{else ret (product xs)} & \coqin{Ret 0} \\
\inbra{redundant conditional} & \inbra{\coqin{move <-}}\\
\hask{ret (product xs)} & \coqin{Ret (product s)} \\
\end{tabular}
\caption{Comparison between a paper proof and a proof using \monae{}}
\label{fig:fastprod}
\end{figure}

Figure~\ref{fig:fastprod} displays the pencil-and-paper proof and the
\coq{} proof that \coqin{fastprod} is pure, i.e., that it never throws
an unhandled exception. Both proofs are essentially the same, though
in practice the \coq{} proof will be streamlined in a script of two
lines (of less than 80 characters):
\begin{minted}[bgcolor=mygray]{ssr}
Lemma fastprodE s : fastprod s = Ret (product s).
Proof.
rewrite /fastprod /work lift_if if_ext catchfailm.
by rewrite catchret; case: ifPn => // /product0 <-.
Qed.
\end{minted}
The fact that we achieve the same conciseness as the pencil-and-paper
proof is not because the example is simple: the same can be said of
all the examples we mechanized.

\paragraph*{Outline}
In Sect.~\ref{sec:hier}, we explain how we formalize an extensible
hierarchy of interfaces of effects in \coq.
This hierarchy allows for effects to be combined and is the main
ingredient of monadic equational reasoning.
In Sect.~\ref{sec:model_of_monads}, we explain how we provide concrete
models for monads, thus showing that the equations defining an
effects are consistent.
This includes in particular a formalization of monad transformers in
Sect.~\ref{sec:monad_transformer}.
To illustrate the use of this framework, we explain in
Sect.~\ref{sec:quicksort} the formalization of an existing program
derivation by \cite{mu2020flops}; we focus on the setting of this
experiment instead of the step-by-step program derivation.
Section \ref{sec:typed_store} demonstrates that we can use \monae{} to
investigate the design of new equational theories; concretely, we
propose an equational theory for the \newterm{typed store monad}
allowing for reasoning on ML-like languages such as \ocaml.
We review related work in Sect.~\ref{sec:related_work} and conclude
in Sect.~\ref{sec:conclusion}.


\section{An extensible implementation of interfaces of effects}
\label{sec:hier}

\subsection{\hb{} in a nutshell}
\label{sec:hb}

\hb{} extends \coq{} with commands to define hierarchies of
mathematical structures and is used in the Mathematical Components (hereafter, \mathcomp) library~\citep{mathcompbook}.
Each mathematical structure is presented as
three layers: a carrier, operations, and properties.  This layering is
standard in mathematics\footnote{This layering is also referred to as
  the ``stuff, structure, and property'' principle in a modern
  context~\citep[Sect.~2.4]{baez2010}.}, thus enabling a
straightforward translation of the textbook definition of a structure
in many usual applications.

The commands are designed so that
hierarchies can evolve (for example by splitting a structure into
smaller structures) without breaking existing code.
In addition to commands to build hierarchies, \hb{} also checks their
validity by detecting missing interfaces~\citep[Sect.~6]{saito2022mpc}
or competing inheritance paths~\citep{affeldt2020ijcar}.

The main concept of \hb{} is the one of \newterm{factory}. This is a
record defined by the command \coqin{HB.factory} that packs
operations and properties for a carrier, providing an interface
to the theory of the mathematical structures that incorporate the factory.

\newterm{Mixins} (defined by the command \coqin{HB.mixin}) are
the basic factories that list operations and properties to be
included in the definition of a mathematical structure.

\newterm{Structures} (defined by the command \coqin{HB.structure})
pack a carrier with one or more factories (intuitively, forming a sigma-type)
to complete them into a mathematical structure.
More precisely, the command
\begin{minted}[bgcolor=mygray]{ssr}
HB.structure Definition M := {A of f1 & f2 & ... & fn}.
\end{minted}
equips \coqin{A} with the interfaces (factories) \coqin{f1}, \coqin{f2}, \ldots,
\coqin{fn}.  Concretely, it creates a \coq{} \coqin{Record} with a
parameter corresponding to \coqin{A} in which each field is one of the
interfaces applied to the parameter, a dependent pair whose first
field corresponds to \coqin{A} and whose second field is an instance
of the record mentioned just above, a module that contains the
\coqin{Record} and the dependent pair, and unification
hints~\citep[Sect.~3.2]{cohen2020fscd}.
In fine, \hb{} commands compile mathematical structures to
\newterm{packed classes}~\citep{garillot2009tphols} but the technical
details (\coq{} modules, records, coercions, implicit arguments,
canonical structures instances, notations, etc.) are hidden to the
user.

Factories (including mixins) can be instantiated (command
\coqin{HB.instance}) with concrete objects. Instances are built with
\coqin{.Build} functions that are automatically generated for each
factory.

A \newterm{builder} is a function that shows by construction that a factory is
sufficient to build a mixin. To write a builder, one uses the command
\coqin{HB.builders} that opens a \coq{} section starting from a
factory and ending with instances of mixins.

\subsection{Functors and natural transformations}
\label{sec:functor_nattrans}

Our hierarchy of interfaces of effects starts with capturing the
notion of functors on the category of sets.  In our \coq{} definition,
the domain and codomain of functors are fixed to \coq's native type
\coqin{Type}: this a predicative type that can be interpreted as the
universe of sets in set-theoretic semantics.  Hereafter, \coqin{UU0}
is an alias for \coqin{Type}\footnote{We introduce the alias
  \coqin{UU0} for \coqin{Type} for practical reasons: although a
  predicative \coqin{UU0} is appropriate for the results presented
  here, there is at least one application~\citep{affeldt2020types}
  that requires \coqin{UU0} to be impredicative. The \coqin{UU0} alias
  provides an easy way to substitute \coq's \coqin{Type} for \coq's
  \coqin{Set} which is impredicative without modifying other parts of
  our development.}.

Using \hb{}, we define functors by the mixin \coqin{isFunctor} (line
\ref{line:isFunctor}).  The carrier is a function \coqin{F} of type
\coqin{UU0 -> UU0} (line \ref{line:isFunctor}) that represents the
action on objects and the operator \coqin{actm} (line \ref{line:actm})
represents the action on morphisms.
The functor laws appear in lines \ref{line:functorid} and
\ref{line:functorcomp} (they correspond to the standard definitions,
see Table~\ref{tab:laws_monae} for details).
\begin{minted}[bgcolor=mygray,numbers=left,xleftmargin=1.5em,escapeinside=88]{ssr}
HB.mixin Record isFunctor (F : UU0 -> UU0) := { 8\label{line:isFunctor}8
  actm : forall A B : UU0, (A -> B) -> F A -> F B ; 8\label{line:actm}8
  functor_id : FunctorLaws.id actm ; 8\label{line:functorid}8
  functor_o : FunctorLaws.comp actm 8\label{line:functorcomp}8 }.
\end{minted}
Given a functor \coqin{F} and a \coq{} function \coqin{f} (seen as a
morphism), we denote by \mintinline{ssr}{F # f}
the action of~\coqin{F} on~\coqin{f}.

We can now create instances of the type \coqin{functor}. For example,
we can equip \coqin{idfun}, the standard identity function of \coq,
with the structure of \coqin{functor} by using the \coqin{HB.instance}
command (line \ref{line:functor_idfun} below). It is essentially a
matter of proving that the functor laws are trivially satisfied (lines
\ref{line:functor_idfun_id}, \ref{line:functor_idfun_comp}):
\begin{minted}[bgcolor=mygray,numbers=left,xleftmargin=1.5em,escapeinside=88]{ssr}
Section functorid.
Let id_actm (A B : UU0) (f : A -> B) : idfun A -> idfun B := f.
Let id_id : FunctorLaws.id id_actm. Proof. by []. Qed. 8\label{line:functor_idfun_id}8
Let id_comp : FunctorLaws.comp id_actm. Proof. by []. Qed. 8\label{line:functor_idfun_comp}8
HB.instance Definition _ := isFunctor.Build idfun id_id id_comp. 8\label{line:functor_idfun}8
End functorid.
\end{minted}
Similarly, we define an instance for the \coq{} composition of function (notation \coqin{\o}).
\begin{minted}[bgcolor=mygray,numbers=left,xleftmargin=1.5em,escapeinside=88]{ssr}
Section functor_composition.
Variables f g : functor.
Let comp_actm (A B : UU0) (h : A -> B) : (f \o g) A -> (f \o g) B :=
  f # (g # h).
Let comp_id : FunctorLaws.id comp_actm.
Proof. (* use functor_id twice *) Qed.
Let comp_comp : FunctorLaws.comp comp_actm.
Proof. (* use functor_o twice and functional extensionality *) Qed.
HB.instance Definition _ := isFunctor.Build (f \o g) comp_id comp_comp.
End functor_composition.
\end{minted}

\def\myop{\,{\small\texttt{op}}\,}
\def\jointt{{\small\texttt{join}}}
\def\rettt{{\small\texttt{ret}}}
\def\myid{\textsl{id}}
\def\bindop#1#2{#1 \gg\!= #2}
\def\bindseq#1#2{#1 \gg #2}

\begin{table}
\centering

\caption{Algebraic laws used in this paper.
See \cite[file \coqin{hierarchy.v}]{monae} for the code.}
\label{tab:laws_monae}
\begin{tabular}{|l|l|}
\hline
\multicolumn{2}{|l|}{\coqin{Module FunctorLaws.}} \\
\hline
\coqin{id f} & $\coqin{f} \, \myid = \myid$ \\
\coqin{comp f} & $\coqin{f}\, (g \circ h) = \coqin{f}\, g \circ \coqin{f}\, h$ \\
\hline
\multicolumn{2}{|l|}{\coqin{Module JoinLaws.} (given a functor \coqin{F})} \\
\hline
\coqin{right_unit ret join} & $\jointt \circ (\coqin{F}\ \rettt) = \myid$ \\
\coqin{left_unit ret join} & $\jointt \circ \rettt = \myid$ \\
& {\footnotesize (where $\rettt$ has \coqin{F} as an implicit parameter)} \\
\coqin{associativity join} & $\jointt \circ (\coqin{F}\ \jointt) = \jointt \circ \jointt$ \\
& {\footnotesize (where \coqin{F} is an implicit parameter of the right-most $\jointt$)} \\
\hline
\multicolumn{2}{|l|}{\coqin{Module BindLaws.} (where $\bindop{\cdot}{\cdot}$ is a notation for the identifier \coqin{bind})} \\
\hline
\coqin{associative bind} & $\bindop{(\bindop{m}{f})}{g} = \bindop{m}{\lambda x.(\bindop{f(x)}{g})}$ \\
\coqin{left_id op ret} & $\rettt\myop m=m $ \\
\coqin{right_id op ret} & $m\myop \rettt=m $ \\
\coqin{left_neutral bind ret} & $\bindop{\rettt}{f} = f$ \\
\coqin{right_neutral bind ret} & $\bindop{m}{\rettt} = m$ \\
\coqin{left_zero bind z} & $\bindop{\coqin{z}}{f} = \coqin{z}$ \\
\coqin{right_zero bind z} & $\bindop{m}{\coqin{z}} = \coqin{z}$\\
\coqin{left_distributive bind op} & $\bindop{m\myop n}{f} = (\bindop{m}{f}) \myop (\bindop{n}{f})$ \\
\coqin{right_distributive bind op} & $\bindop{m}{\lambda x.(f\,x)\myop(g\,x)} = (\bindop{m}{f}) \myop (\bindop{m}{g})$ \\
\hline
\end{tabular}
\end{table}

As explained in Sect.~\ref{sec:hb},
the mixin \coqin{isFunctor} only provides an interface, the type of functors
is defined by a \hb{} structure \coqin{Functor} that can be
seen as a sigma-type of a carrier \coqin{F} satisfying the
interface \coqin{isFunctor}:
\begin{minted}[bgcolor=mygray]{ssr}
#[short(type=functor)]
HB.structure Definition Functor := {F of isFunctor F}.
\end{minted}

We now define natural transformations. Given two functors \coqin{F}
and \coqin{G}, we formalize the components of natural
transformations as a family of functions \coqin{f}, of type \coqin{forall A, F A -> G A}
(we note this type \coqin{F ~~> G} for short) that satisfies the following predicate
(recall that \coqin{\o} denotes function composition):
\begin{minted}[bgcolor=mygray]{ssr}
Definition naturality (F G : functor) (f : F ~~> G) :=
  forall (A B : UU0) (h : A -> B), (G # h) \o f A = f B \o (F # h).
\end{minted}

Natural transformations are defined by means of the mixin and the
structure below.
\begin{minted}[bgcolor=mygray]{ssr}
HB.mixin Record isNatural (F G : functor) (f : F ~~> G) := {
  natural : naturality F G f }.

#[short(type=nattrans)]
HB.structure Definition Nattrans (F G : functor) :=
  {f of isNatural F G f}.
\end{minted}
In \monae{}, we note \coqin{F ~> G} instead of \coqin{nattrans F G} the natural transformation from \coqin{F} to \coqin{G}.

\subsection{Formalization of monads}
\label{sec:formalization_monads}

We now formalize monads. A monad extends a functor with two natural transformations:
the unit \coqin{ret} (line \ref{line:isMonad_ret} below) and the multiplication \coqin{join} (line \ref{line:isMonad_join}).
They satisfy three laws (lines \ref{line:joinretM}--\ref{line:joinA}, see Table~\ref{tab:laws_monae}).
Furthermore, we add to the mixin an identifier for the \coqin{bind}
operator (line \ref{line:isMonad_bind}, hereafter noted \coqin{>>=} or
\coqin{>>} when the right hand side ignores its input) and an equation
that defines \coqin{bind} in term of unit and multiplication (line
\ref{line:bindE}). Note however that this does not mean that the
creation of a new instance of monads requires the (redundant)
definition of the unit, multiplication, and \coqin{bind} (this is
explained below).

\begin{minted}[bgcolor=mygray,numbers=left,xleftmargin=1.5em,escapeinside=88]{ssr}
HB.mixin Record isMonad (F : UU0 -> UU0) of Functor F := { 8\label{line:monad_is_functor}8
  ret : idfun ~> F ; 8\label{line:isMonad_ret}8
  join : F \o F ~> F ; 8\label{line:isMonad_join}8
  bind : forall (A B : UU0), F A -> (A -> F B) -> F B ; 8\label{line:isMonad_bind}8
  bindE : forall (A B : UU0) (f : A -> F B) (m : F A),
    bind A B m f = join B ((F # f) m) ; 8\label{line:bindE}8
  joinretM : JoinLaws.left_unit ret join ; 8\label{line:joinretM}8
  joinMret : JoinLaws.right_unit ret join ; 8\label{line:joinMret}8
  joinA : JoinLaws.associativity join }. 8\label{line:joinA}8

#[short(type=monad)]
HB.structure Definition Monad := {F of isMonad F &}. 8\label{line:monad_structure}8
\end{minted}
The fact that a monad extends a functor can be observed at line
\ref{line:monad_is_functor} with the \coqin{of} keyword.  Also, when
declaring the structure at line \ref{line:monad_structure}, the
\coqin{&} mark indicates inheritance w.r.t.\ all the mixins on which
the structure depends on \footnote{\monae{} also features a
  formalization of (concrete) categories that has been used to
  formalize the geometrically convex monad
  \citep[Sect.~5]{affeldt2021jfp}. Both are connected in the sense
  that a monad over the category corresponding to the type
  \coqin{Type} of \coq{} (seen as a Grothendieck universe) can be used
  to instantiate the \coqin{isMonad} interface. Yet, as far as this
  paper is concerned, this generality is not useful.}.  Hereafter, we
also use the notation \coqin{Ret} instead of \coqin{ret}; this is for
technical reasons that have to do with the setting of implicit
arguments.

The above definition of monads is not the privileged interface to define new
instances of monads. We also provide factories with a smaller interface from
which the above mixin is recovered. For example, here is the factory to build
monads from the unit and the multiplication:
\begin{minted}[bgcolor=mygray]{ssr}
HB.factory Record isMonad_ret_join (F : UU0 -> UU0) of isFunctor F := {
  ret : idfun ~> F ;
  join : F \o F ~> F ;
  joinretM : JoinLaws.left_unit ret join ;
  joinMret : JoinLaws.right_unit ret join ;
  joinA : JoinLaws.associativity join }.
\end{minted}
This corresponds to the textbook definition of a monad, since it does
not require the simultaneous definition of the unit, the
multiplication, and \coqin{bind}. We use the \coqin{HB.builders}
command (Sect. \ref{sec:hb}) to show that this lighter definition is
sufficient to satisfy the \coqin{isMonad} interface.

Similarly, there is a factory to build monads from the unit and \coqin{bind} only:
\begin{minted}[bgcolor=mygray]{ssr}
HB.factory Record isMonad_ret_bind (F : UU0 -> UU0) := {
  ret : forall (A : UU0), A -> F A ;
  bind : forall (A B : UU0), F A -> (A -> F B) -> F B ;
  bindretf : BindLaws.left_neutral bind ret ;
  bindmret : BindLaws.right_neutral bind ret ;
  bindA : BindLaws.associative bind }.  
\end{minted}

The definition of monad (interface \coqin{isMonad}) that we presented
here is an improvement compared to the original formalization
\citep[Sect. 2.1]{affeldt2019mpc} because there is now an explicit
type of natural transformations (for \coqin{ret} and \coqin{join}) and
because \hb{} guarantees that monads instantiated by factories do
correspond to the same type monad.  See \cite[file
\coqin{monad_model.v}]{monae} for many instances of the monad
structure handled by the \coqin{isMonad_ret_bind} factory.

\subsection{Nondeterminism monad}
\label{sec:nondeterminism}

In the previous section, we explained the case of a simple extension:
one structure (the one of monads) that extends another (the one of
functors). In this section we explain how a structure combines several
interfaces.

The nondeterminism monad extends both the failure monad and the choice
monad. The failure monad \coqin{failMonad} extends the class of monads
(Sect.~\ref{sec:formalization_monads}) with a failure operator
\coqin{fail} (line~\ref{line:fail} below) that is a left-zero of
\coqin{bind} (line~\ref{line:bindfailf}). This is the same extension
methodology as in Sect.~\ref{sec:formalization_monads}:
\begin{minted}[bgcolor=mygray,numbers=left,xleftmargin=1.5em,escapeinside=88]{ssr}
HB.mixin Record isMonadFail (M : UU0 -> UU0) of Monad M := {
  fail : forall A : UU0, M A ; 8\label{line:fail}8
  bindfailf : BindLaws.left_zero (@bind M) fail }. 8\label{line:bindfailf}8

#[short(type=failMonad)]
HB.structure Definition MonadFail := {M of isMonadFail M & }.
\end{minted}
The prefix \coqin{@} at line \ref{line:bindfailf} is a \coq{} notation
to make all the arguments of a function explicit.

\begin{table}
\centering
\caption{Algebraic laws defined in \mathcomp{}}
\label{tab:laws_ssreflect}
\begin{tabular}{|l|l|}
\hline
\coqin{associative op} & $x\myop(y\myop z) = (x\myop y)\myop z$ \\
\coqin{left_id e op} & $e \myop x=x$ \\
\coqin{right_id e op} & $x \myop e=x$ \\
\coqin{left_zero z op} & $z \myop x=z$ \\
\coqin{idempotent op} & $x \myop x = x$ \\
\hline
\end{tabular}
\end{table}

The choice monad \coqin{altMonad} extends the class of monads with a
choice operator \coqin{alt} (line~\ref{line:alt} below, infix
notation: \coqin{[~]}) that is associative (line~\ref{line:altA}) and
such that \coqin{bind} distributes leftwards over choice
(line~\ref{line:alt_bindDl}):
\begin{minted}[bgcolor=mygray,numbers=left,xleftmargin=1.5em,escapeinside=88]{ssr}
HB.mixin Record isMonadAlt (M : UU0 -> UU0) of Monad M := {
  alt : forall T : UU0, M T -> M T -> M T ; 8\label{line:alt}8
  altA : forall T : UU0, associative (@alt T) ; 8\label{line:altA}8
  alt_bindDl : BindLaws.left_distributive (@bind M) alt }. 8\label{line:alt_bindDl}8

#[short(type=altMonad)]
HB.structure Definition MonadAlt := {M of isMonadAlt M & }.
\end{minted}
See Tables~\ref{tab:laws_monae} and \ref{tab:laws_ssreflect} for the
definition of the predicates \coqin{associative} and
\coqin{left_distributive} respectively.

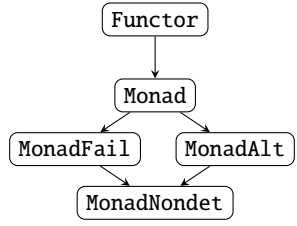
\begin{figure}
\centering
\begin{tikzpicture}
\begin{scope}[every node/.style={draw,fill=white,rounded corners=0.3em},
              every path/.style={-stealth}]

\node (functor)  {\coqin{Functor}} ;
\node (monad) [below of=functor] {\coqin{Monad}} ;
\node (fail) [below left of=monad,xshift=-1em] {\coqin{MonadFail}} ;
\node (alt) [below right of=monad,xshift=1em] {\coqin{MonadAlt}} ;
\node (nondet) [below right of=fail,xshift=1em] {\coqin{MonadNondet}} ;

\draw (functor) -- (monad) ;
\draw (monad) -- (fail) ;
\draw (monad) -- (alt) ;
\draw (fail) -- (nondet) ;
\draw (alt) -- (nondet) ;
\end{scope}
\end{tikzpicture}
\caption{Hierarchy of effects for the nondeterminism monad}
\label{fig:hier_nondet}
\end{figure}

The nondeterminism monad \coqin{nondetMonad} defined below extends
both the failure monad and the choice monad (as can be seen at
line~\ref{line:nondet_inherit} below, see also
Fig.~\ref{fig:hier_nondet}) and adds laws expressing that failure is
an identity of choice (lines \ref{line:altfailm}, \ref{line:altmfail})
\begin{minted}[bgcolor=mygray,numbers=left,xleftmargin=1.5em,escapeinside=88]{ssr}
HB.mixin Record isMonadNondet (M : UU0 -> UU0)
    of MonadFail M & MonadAlt M := { 8\label{line:nondet_inherit}8
  altfailm : @BindLaws.left_id M (@fail M) (@alt M) ; 8\label{line:altfailm}8
  altmfail : @BindLaws.right_id M (@fail M) (@alt M) }. 8\label{line:altmfail}8

#[short(type=nondetMonad)]
HB.structure Definition MonadNondet :=
  { M of isMonadNondet M & MonadFail M & MonadAlt M }.
\end{minted}
As a result, a nondeterminism monad can be regarded both as a failure
monad or as a choice monad.

\subsubsection{Assertions in monadic programs}
\label{sec:dassert}

The introduction of the failure monad already allows for the
definition of reusable library functions.

Given a failure monad \coqin{M}, it is customary to define assertions
as follows. A computation \coqin{guard b} of type \coqin{M unit} fails
or skips according to a boolean value \coqin{b}:
\begin{minted}[bgcolor=mygray]{ssr}
Definition guard (b : bool) : M unit := locked (if b then skip else fail).
\end{minted}

An assertion \coqin{assert p a} is a computation of type \coqin{M A}
that fails or returns \coqin{a} according to whether \coqin{p a} is
true or not (\coqin{pred} is the type of boolean predicates in \mathcomp):
\begin{minted}[bgcolor=mygray]{ssr}
Definition assert {A : UU0} (p : pred A) (a : A) : M A :=
  locked (guard (p a) >> Ret a).
\end{minted}

It turns out that in practice it is even useful to define a
\newterm{dependently-typed assertion} that fails or returns a value
{\em together with a proof\/} that the predicate is satisfied:
\begin{minted}[bgcolor=mygray]{ssr}
Definition dassert (p : pred A) a : M { a | p a } :=
  if Bool.bool_dec (p a) true is left pa then Ret (exist _ _ pa) else fail.
\end{minted}
The notation \coqin{{a | p a}} is for a dependent pair whose constructor
is \coqin{exist}. To demonstrate the usefulness of dependently-typed
assertions we need a bit of context that will be provided by a
concrete example in Sect.~\ref{sec:qperm}.

\subsection{Exception monad} 
\label{sec:exceptMonad}

Similarly to the nondeterminism monad that extends the failure monad
in the previous section, the exception monad can be defined by
extending the failure monad with a \coqin{catch} operator and four
laws:
\begin{minted}[bgcolor=mygray]{ssr}
HB.mixin Record isMonadExcept (M : UU0 -> UU0) of MonadFail M := {
  catch : forall A, M A -> M A -> M A ;
  catchmfail : forall A, right_id fail (@catch A) ;
  catchfailm : forall A, left_id fail (@catch A) ;
  catchA : forall A, associative (@catch A) ;
  catchret : forall A x, @left_zero (M A) (M A) (Ret x) (@catch A) }.

#[short(type=exceptMonad)]
HB.structure Definition MonadExcept := {M of isMonadExcept M & }.
\end{minted}
See Table~\ref{tab:laws_ssreflect} for the definitions of the generic
predicates used in this interface. As for an application, we send back
the reader to the \coqin{fastprod} example explained in
Fig.~\ref{fig:fastprod} (Sect.~\ref{sec:introduction}).

\subsection{State monad}
\label{sec:stateMonad}

The state monad is certainly the first monad that comes to mind when
speaking of effects. It denotes computations that transform a state
(type \coqin{S} below). It comes with a \coqin{get} operator to yield
a copy of the state and a \coqin{put} operator to overwrite it. These
functions are constrained by four laws \citep{gibbons2011icfp}
that appear below at lines \ref{line:putput}--\ref{line:getget}:
\begin{minted}[bgcolor=mygray,numbers=left,xleftmargin=1.5em,escapeinside=88]{ssr}
HB.mixin Record isMonadState (S : UU0) (M : UU0 -> UU0) of Monad M := {
  get : M S ;
  put : S -> M unit ;
  putput : forall s s', put s >> put s' = put s' ; 8\label{line:putput}8
  putget : forall s, put s >> get = put s >> Ret s ;
  getputskip : get >>= put = skip ;
  getget : forall (A : UU0) (k : S -> S -> M A),
    get >>= (fun s => get >>= k s) = get >>= fun s => k s s }. 8\label{line:getget}8

#[short(type=stateMonad)]
HB.structure Definition MonadState (S : UU0) :=
  { M of isMonadState S M & }.  
\end{minted}

\subsection{Array monad}
\label{sec:arrayMonad}

\begin{table}
\caption{The laws of the array monad}
\label{tab:arrayMonad}
\centering
\begin{tabular}{|l|l|}
\hline
\coqin{aputput}   & \coqin{aput i v >> aput i v' = aput i v'} \\
\coqin{aputget}   & \coqin{aput i v >> aget i >>= k = aput i v >> k v} \\
\coqin{agetputskip} & \coqin{aget i >>= aput i = skip} \\
\coqin{agetget} & \coqin{aget i >>= (fun v => aget i >>= k v) =} \\
             & \coqin{aget i >>= fun v => k v v} \\
\coqin{agetC}   & \coqin{aget i >>= (fun u => aget j >>= (fun v => k u v)) =} \\
             & \coqin{aget j >>= (fun v => aget i >>= (fun u => k u v))} \\
\coqin{aputC}   & \coqin{i != j \/ u = v ->} \\
             & \coqin{  aput i u >> aput j v = aput j v >> aput i u} \\
\coqin{aputgetC}  & \coqin{i != j ->} \\
               & \coqin{  aput i u >> aget j >>= k =} \\
               & \coqin{  aget j >>= (fun v => aput i u >> k v)} \\
\hline
\end{tabular}
\end{table}

The array monad extends a basic monad with a notion of indexed array
\citep[Sect.~5.1]{mu2020flops}. Intuitively, it is a
generalization of the state monad (Sect.~\ref{sec:stateMonad}).  It
provides two operators to read and write indexed cells.  Given an
index \coqin{i}, \coqin{aget i} returns the value stored at \coqin{i}
and \coqin{aput i v} stores the value \coqin{v} at \coqin{i}.  These
operators satisfy the laws of Table~\ref{tab:arrayMonad}.
For example, \coqin{aputput} means that the result of storing the
value~\coqin{v} at index \coqin{i} and then storing the value~\coqin{v'} at
index~\coqin{i} is the same as the result of storing the value~\coqin{v'}.
The law \coqin{aputget} means that it is not necessary to get a value
after having stored it provided this value is directly passed to the
continuation. Other laws can be interpreted similarly.

In \monae{}, the array monad is formalized by extending the base monad
with the following mixin:
\begin{minted}[bgcolor=mygray,escapeinside=88]{ssr}
HB.mixin Record isMonadArray (S : UU0) (I : eqType) (M : UU0 -> UU0)
    of Monad M := {
  aget : I -> M S ;
  aput : I -> S -> M unit ;
  aputput : forall i s s', aput i s >> aput i s' = aput i s' ;
  (* other laws similarly mimics the laws of 8Table \ref{tab:arrayMonad}8,
     see 8\cite[file {\tt hierarchy.v}]{monae}8 for implementation details *)
  }.

#[short(type=arrayMonad)]
HB.structure Definition MonadArray (S : UU0) (I : eqType) :=
  { M of isMonadArray S I M & isMonad M & isFunctor M }.
\end{minted}
Note that the type of indices is an \coqin{eqType}, i.e., a type with
decidable equality, as required by the laws of the array monad.

Using the array monad, we can write various functions that manipulate arrays.
For example, we can swap the contents of two cells by combining
\coqin{aget} and \coqin{aput}:
\begin{minted}[bgcolor=mygray,escapeinside=88]{ssr}
Definition aswap i j : M unit :=
  aget i >>= (fun x => aget j >>= (fun y => aput i y >> aput j x)).
\end{minted}
We can also recursively call the \coqin{aput} operator to write a list
\coqin{xs} to the array starting from the index \coqin{i}:
\begin{minted}[bgcolor=mygray]{ssr}
Fixpoint writeList i (s : seq E) : M unit :=
  if s isn't x :: xs then Ret tt else aput i x >> writeList i.+1 xs.
\end{minted}

\subsection{Probability monad}
\label{sec:probMonad}

\def\pchoiceleft{\triangleleft}
\def\pchoiceright{\triangleright}
\def\pchoice#1#2#3{#1 \pchoiceleft #2 \pchoiceright #3}

Before defining the probability monad, we define a type \coqin{{prob R}} with \coqin{R} a type for real numbers to
represent ``probabilities'', i.e., real numbers between $0$ and
$1$. This definition comes with a notation \mintinline{ssr}{p
such that the type \coqin{{prob R}} is automatically inferred based on the shape of
the expresson~\coqin{p}. For example, \mintinline{ssr}{1
represents the real number~$1$ seen as a probability.

\def\sof#1#2{\textrm{s}(#1,#2)}
\def\rof#1#2{\textrm{r}(#1,#2)}

In order to define the interface of the probability monad in a modular way, we introduce
as an intermediate step the interfaces of ``convex monads'', i.e., an
interface that provides a family of binary \coqin{choice}
$\pchoice{\cdot}{p}{\cdot}$ operators parameterized by probabilities
$p$ (see line~\ref{line:choice} below).  These choice operators
satisfy the axioms of \newterm{convex spaces}~\cite[Def~3]{jacobs2010tcs}:
\begin{itemize}
\item $\pchoice{a}{1}{b} = 1$ (line~\ref{line:choice1}),
\item $\pchoice{a}{p}{b} = \pchoice{b}{\bar{p}}{a}$ (skewed commutativity, line~\ref{line:choiceC}),
\item $\pchoice{a}{p}{a} = a$ (idempotence, line~\ref{line:choicemm}), and
\item $\pchoice{a}{p}{\pchoice{b}{q}{c}} = \pchoice{\pchoice{a}{\sof{p}{q}}{b}}{\rof{p}{q}}{c}$
where $\sof{p}{q} \isdef \overline{\bar{p}\bar{q}}$
and $\rof{p}{q} \isdef \frac{p}{\sof{p}{q}}$ (quasi associativity, line~\ref{line:choiceA}).
\end{itemize}
\begin{minted}[bgcolor=mygray,numbers=left,xleftmargin=1.5em,escapeinside=88]{ssr}
HB.mixin Record isMonadConvex {R : realType} (M : UU0 -> UU0)
    of Monad M := {
  choice : forall (p : {prob R}) (T : UU0), M T -> M T -> M T ; 8\label{line:choice}8
  choice1 : forall (T : UU0) (a b : M T), choice 1%:pr _ a b = a ; 8\label{line:choice1}8
  choiceC : forall (T : UU0) p (a b : M T),
    choice p _ a b = choice (p.~%:pr) _ b a ; 8\label{line:choiceC}8
  choicemm : forall (T : UU0) p, idempotent (@choice p T) ; 8\label{line:choicemm}8
  choiceA : forall (T : UU0) (p q r s : {prob R}) (a b c : M T), 8\label{line:choiceA}8
    choice p _ a (choice q _ b c) =
    choice [s_of p, q] _ (choice [r_of p, q] _ a b) c }. 
\end{minted}
Note that at line \ref{line:choiceC}, the notation \coqin{r.~} stands for $\bar{r} \isdef 1 - r$.

The probability monad merely extends the convex monad by adding the axiom that
\coqin{bind} left-distributes over probabilistic choice (line~\ref{line:choice_bindDl}):
\begin{minted}[bgcolor=mygray,numbers=left,xleftmargin=1.5em,escapeinside=88]{ssr}
HB.mixin Record isMonadProb {R : realType} (M : UU0 -> UU0)
    of MonadConvex R M := {
  choice_bindDl : forall p, 8\label{line:choice_bindDl}8
    BindLaws.left_distributive (@bind M) (choice p) }. 

#[short(type=probMonad)]
HB.structure Definition MonadProb {R : realType} :=
  {M of isMonadProb R M & }.
\end{minted}

\subsubsection{Probability and nondeterminism}
\label{sec:altprob_monad}

Last, we briefly mention the interface of a monad that mixes probability and
nondeterminism: the \newterm{geometrically convex monad}~\citep{cheungPhD2017}.
Its interface just adds the right distributivity of probabilistic choice over
nondeterministic choice to the probability monad:
\begin{minted}[bgcolor=mygray]{ssr}
HB.mixin Record isMonadAltProb {R : realType} (M : UU0 -> UU0)
    of MonadAltCI M & MonadProb R M :=
  { choiceDr : forall T p, right_distributive
      (@choice R M p T) (fun a b => a [~] b) }.

#[short(type=altProbMonad)]
HB.structure Definition MonadAltProb {R : realType} :=
  { M of isMonadAltProb R M & isFunctor M & isMonad M & isMonadAlt M &
         isMonadAltCI M & isMonadProb R M & isMonadConvex R M }.
\end{minted}
The formalization of this monad is the topic of previous
work~\citep{affeldt2021jfp} which explains that the construction of
its model is important but not trivial.

\section{Model of monads}
\label{sec:model_of_monads}

In the previous section (Sect.~\ref{sec:hier}), we explained how we
build a hierarchy of interfaces that allows for combining effects.  In
this section, we explain how we formalize models to show that the
equations in interfaces are indeed consistent. We only consider the
examples of monad transformers and of the probability monad; see
\cite[file \coqin{monad_model.v}]{monae} for more examples, and in particular \cite{affeldt2021jfp}
for the geometrically convex monad.

\subsection{Monad transformers}
\label{sec:monad_transformer}

Monad transformers is a well-known approach to combine monads
\citep{liang1995popl} that is commonly used to write Haskell programs.
The interest in extending \monae{} with monad transformers is twofold:
it provides an application of monadic equational reasoning (for
example, Jaskelioff's theory of modular monad transformers has been
formalized in \monae{} \citep{affeldt2020types}) and monad
transformers can be used to formalize concrete models of monads
(as we will see in Sect.~\ref{sec:typed_store_model}).

\begin{table}
\centering
\caption{Algebraic laws for monad morphisms}
\label{tab:laws_monadM}
\begin{tabular}{|l|l|}
\hline
\multicolumn{2}{|l|}{\coqin{Module MonadMLaws.} (where \coqin{e} has type \coqin{M ~~> N})} \\
\hline
\coqin{ret e} & \coqin{forall A : UU0, e A \o Ret = Ret} \\
\coqin{bind e} & \coqin{forall (A B : UU0) (m : M A) (f : A -> M B),}\\
& \coqin{  e B (m >>= f) = e A m >>= (e B \o f)} \\
\hline
\end{tabular}
\end{table}

Similarly to functors, monads, and interface for effects, monad
transformers can be formalized in terms of \hb{} interfaces and
structures.
Given two monads \coqin{M} and \coqin{N}, a monad morphism is a
function \coqin{M ~~> N} (see Sect.~\ref{sec:functor_nattrans} for the
definition of \coqin{~~>}) that satisfies the laws of monad
morphisms~\citep[Def.~19]{benton2000appsem}
\citep[Def.~7]{jaskelioff2009esop}:
\begin{minted}[bgcolor=mygray]{ssr}
HB.mixin Record isMonadM (M N : monad) (e : M ~~> N) := {
  monadMret : MonadMLaws.ret e ;
  monadMbind : MonadMLaws.bind e }.
\end{minted}
Table~\ref{tab:laws_monadM} provides the definitions of the laws.  An important property
of monad morphisms is that they are natural transformations, so that
we define the type of monad morphisms using the interface of monad
morphisms {\em and\/} the interface of natural transformations (Sect.~\ref{sec:functor_nattrans}):
\begin{minted}[bgcolor=mygray]{ssr}
#[short(type=monadM)]
HB.structure Definition MonadM (M N : monad) :=
  {e of isMonadM M N e & isNatural M N e}.  
\end{minted}
However, since the laws of monad morphisms imply naturality, a monad
morphism can be defined directly by a factory with {\em only} the laws of
monad morphisms:
\begin{minted}[bgcolor=mygray]{ssr}
HB.factory Record isMonadM_ret_bind (M N : monad) (e : M ~~> N) := {
  monadMret : MonadMLaws.ret e ;
  monadMbind : MonadMLaws.bind e }.
\end{minted}
Upon declaration of this factory, we can use the \coqin{HB.builders} command of \hb{}
to declare instances of the monad morphism and of the natural transformation interfaces\footnote{It would have been less convoluted to define the \coqin{isMonadM} interface
as inheriting from the \coqin{isNatural} interface but \hb{} does not support yet natural transformation components as carrier, this is why we redeclare the \coqin{isMonadM} interface as a factory and use \coqin{HB.builders} here.}.
Like for monads in Sect.~\ref{sec:formalization_monads}, we are again
in the situation where the textbook definition ought better be sought
in factories rather than in the mixins used to formally define it in the first place.

A monad transformer \coqin{t} is a function from \coqin{monad} to
\coqin{monad} such that for any monad \coqin{M} it returns a monad
morphism from \coqin{M} to \coqin{t M}
\citep[Sect.~3.3]{benton2000appsem}
\citep[Def.~9]{jaskelioff2009esop}:
\begin{minted}[bgcolor=mygray]{ssr}
HB.mixin Record isMonadT (T : monad -> monad) := {
  Lift : forall M, monadM M (T M) }.

#[short(type=monadT)]
HB.structure Definition MonadT := {T of isMonadT T}.  
\end{minted}
Going one step further, we can define \newterm{functorial monad
  transformers} \citep[Def.~20]{jaskelioff2009esop} and revise the
formalization of Jaskelioff's modular monad transformers using
\hb{} following \cite{affeldt2020types}, see \citep{monae} for details.

\subsubsection{Example: The state monad transformer}
\label{sec:stateT}

We explain how we instance the type of monad transformer with the
example of the state monad transformer.
Let us assume some type \coqin{S : UU0} for states and some
monad~\coqin{M}.  First, we define the action on objects of the monad
transformed by the state monad transformer:
\begin{minted}[bgcolor=mygray]{ssr}
Definition MS := fun A : UU0 => S -> M (A * S).
\end{minted}

We also define the unit and the \coqin{bind}
operator of the transformed monad:
\begin{minted}[bgcolor=mygray]{ssr}
Definition retS : idfun ~~> MS := fun A : UU0 => curry Ret.
Definition bindS (A B : UU0) (m : MS A) f : MS B :=
  fun s => m s >>= uncurry f.
\end{minted}

Second, we define the monad morphism that will be returned by the lift
operator of the monad transformer.  In \coq{}, we can formalize the
corresponding function by constructing the desired monad assuming~\coqin{M}
(in a \coq{} \coqin{Section}) and using the factory \coqin{isMonad_ret_bind}
from Sect.~\ref{sec:formalization_monads}. It suffices to prove the laws of the monad:
\begin{minted}[bgcolor=mygray]{ssr}
Let bindSretf : BindLaws.left_neutral bindS retS. Proof. ... Qed.
Let bindSmret : BindLaws.right_neutral bindS retS. Proof. ... Qed.
Let bindSA : BindLaws.associative bindS. Proof. ... Qed.
HB.instance Definition _ :=
  isMonad_ret_bind.Build MS bindSretf bindSmret bindSA.
\end{minted}

Then we define the lift operation as a function that given a computation \coqin{m : M A}
returns a computation in the monad \coqin{MS}:
\begin{minted}[bgcolor=mygray]{ssr}
Definition liftS (A : UU0) (m : M A) : MS A :=
  fun s => m >>= (fun x => Ret (x, s)).
\end{minted}
We can finally prove the lift operations satisfy the monad morphism laws of Table~\ref{tab:laws_monadM}:
\begin{minted}[bgcolor=mygray]{ssr}
Let retliftS : MonadMLaws.ret liftS. Proof. ... Qed.
Let bindliftS : MonadMLaws.bind liftS. Proof. ... Qed.
HB.instance Definition _ := isMonadM_ret_bind.Build
  M MS liftS retliftS bindliftS.
\end{minted}
The state monad transformer is defined as an alias \coqin{stateT} to
which the type \coqin{monadT} is associated:
\begin{minted}[bgcolor=mygray]{ssr}
Definition stateT (S : UU0) : monad -> monad := MS S.
HB.instance Definition _ (S : UU0) := isMonadT.Build (stateT S) (@liftS S).
\end{minted}

One should wonder what is the relation between the monads that can be
built with the state monad transformer and the state monad of
Sect.~\ref{sec:stateMonad}. In fact, we can define a \coqin{Put} and a \coqin{Get}
operations for a monad \coqin{MS S M} so that it satisfies the laws of state monads:
\begin{minted}[bgcolor=mygray]{ssr}
Let bindputput s s' : Put s >> Put s' = Put s'. Proof. ... Qed.
Let bindputget s : Put s >> Get = Put s >> Ret s. Proof. ... Qed.
Let bindgetput : Get >>= Put = skip. Proof. ... Qed.
Let bindgetget (A : UU0) (k : S -> S -> stateT S M A) :
  Get >>= (fun s => Get >>= k s) = Get >>= (fun s => k s s).
Proof. ... Qed.
HB.instance Definition _ := @isMonadState.Build
  S (MS S M) Get Put bindputput bindputget bindgetput bindgetget.
\end{minted}
%

\subsection{A model of the probability monad}
As explained in Sect.~\ref{sec:probMonad}, a probability monad has to turn
each input type into a convex space and provide
the \coqin{bind} operator in a way that is compatible with
the convex space structure.
We can achieve these two requirements by, for an input type \coqin{A}, constructing
the type \coqin{{dist A}} of finitely-supported distributions
over \coqin{A}~\citep[Sect.~6.2]{affeldt2019mpc} \citep{infotheo}:
\[
  \coqin{{dist A}} \isdef
  \left\{
    f : \coqin{A} \xlongrightarrow{\text{fin. supp.}} \mathbb R
    \; \middle| \;
    \left(\forall x, 0 < f(x)\right)
    \ \text{and}\ \left(\sum_{x\in\supp(f)} f(x) = 1\right)
  \right\},
\]
\[
  \supp(f) \isdef
  \left\{x \in \coqin{A} \; \middle| \; f(x) \not= 0 \right\}.
\]
Technically, we need \coqin{A} to satisfy the axiom of choice to let
the finitely supported function have a canonical list representation
(appearing as type annotations $\coqin{A : choiceType}$
in the code~\citep{infotheo}).
We assume the generic axiom of choice additionally to \coq{}'s type theory,
and use it to turn every type into a \coqin{choiceType}.
More precisely, the assumption takes the form of a function:
\[
  \coqin{choice_of_Type} : \coqin{Type} \to \coqin{choiceType}.
\]
Composing \coqin{{dist _}} and \coqin{choice_of_Type}, we obtain the
object part of the model $M$ of the probability monad:
\[
  M(A) := \texttt{\{}\coqin{dist}\ (\coqin{choice_of_Type}(A))\texttt{\}}.
\]
%

To complete this definition into a functor, and furthermore, a monad,
we can use the factory \coqin{isMonad_ret_bind} defined in
Sect.~\ref{sec:formalization_monads}, which requires us only to
provide the two monadic operators \coqin{ret} and \coqin{bind},
and to prove that they satisfy the needed laws.

The \coqin{bind} operator computes the weighted sum out of 
a given distribution $p : M(A)$ and a monadic function $g : A \to M(B)$, 
returning a new distribution with probability mass function
\[
  b \longmapsto \sum_{a \in \supp(g)} p(a) \cdot g(a,b).
\]
This function is implemented by the following
code~\citep[file \coqin{fsdist.v}]{infotheo}:
\begin{minted}[bgcolor=mygray]{ssr}
(* Section fsdistbind *)
Variables (A B : choiceType) (p : {dist A}) (g : A -> {dist B}).
Let D := \bigcup_(d <- g @` finsupp p) finsupp d. (* new support for f *)
Let f : {fsfun B -> R with 0} :=          (* 0 being the default value *)
  [fsfun b in D => \sum_(a <- finsupp p) p a * (g a) b | 0].
\end{minted}
The resulting operator can be proved to satisfy the monad laws,
for example, associativity
\begin{minted}[bgcolor=mygray]{ssr}
Lemma fsdistbindA (A B C : choiceType) (m : {dist A}) (f : A -> {dist B})
    (g : B -> {dist C}) :
  (m >>= f) >>= g = m >>= (fun x => f x >>= g).
\end{minted}
follows from the distributivity of multiplication over big sum, and
computation on big unions, both handled by \mathcomp's big operator library.

The \coqin{ret} operator embeds a given point $a : A$ into $M(A)$ as the
distribution concentrated on the singleton $\{a\}$ (a.k.a.\ Dirac's delta),
whose probability mass function is
\[
  x \longmapsto
  \begin{cases} 1 & \text{if $x = a$}, \\ 0 & \text{otherwise}. \end{cases}
\]
The corresponding code in \citep{infotheo} is straightforward:
\begin{minted}[bgcolor=mygray]{ssr}
(* Section fsdist1 *)
Variables (A : choiceType) (a : A).
Let D := [fset a].   (* the singleton set containing only a *)
Let f : {fsfun A -> R with 0} := [fsfun b in D => 1 | 0].
\end{minted}
The other monad laws involving both \coqin{ret} and \coqin{bind} are proved
similarly as above.
\begin{minted}[bgcolor=mygray]{ssr}
Lemma fsdist1bind (A B : choiceType) (a : A) (f : A -> {dist B}) :
  fsdist1 a >>= f = f a.
Lemma fsdistbind1 (A : choiceType) (p : {dist A}) :
  p >>= @fsdist1 A = p.
\end{minted}

We can then pack these definitions and proofs into a monad by applying the
builder function from the factory \coqin{isMonad_ret_bind} as
follows~\citep[file \coqin{proba_monad_model.v}]{monae} :
\begin{minted}[bgcolor=mygray]{ssr}
Definition acto : UU0 -> UU0 := fun A => {dist (choice_of_Type A)}.
Let left_neutral : BindLaws.left_neutral bind ret.
Proof. (* just use fsdist1bind *) Qed.
Let right_neutral : BindLaws.right_neutral bind ret.
Proof. (* fsdistbind1 *) Qed.
Let associative : BindLaws.associative bind.
Proof. (* fsdistbindA *) Qed.
HB.instance Definition _ := isMonad_ret_bind.Build
  acto left_neutral right_neutral associative.
\end{minted}

At this point, we have completed the monad structure
on our construction.  Since finitely supported distributions carry
the convex space structure (thus easily endowed with a
\emph{convex monad} (Sect.~\ref{sec:probMonad}) structure),
the remaining task is to prove the left-distributivity law of \coqin{bind}
over the probabilistic choice:
\begin{minted}[bgcolor=mygray]{ssr}
(* in infotheo *)
Lemma fsdist_conv_bind_left_distr     
  (A B : choiceType) (p : {prob R}) (a b : {dist A}) (f : A -> {dist B}) :
  (a <| p |> b) >>= f  =  (a >>= f) <| p |> (b >>= f).
Proof. (* proved in 8 lines *) Qed.

(* in monae *)
Let prob_bindDl p :                   
  BindLaws.left_distributive (@hierarchy.bind acto) (choice p).
Proof. (* just use the infotheo lemma *) Qed.
\end{minted}

Packing this additional low using the mixin \coqin{isMonadProb},
we finally complete the model of the probability monad:
\begin{minted}[bgcolor=mygray]{ssr}
HB.instance Definition _ := isMonadProb.Build real_realType
  acto prob_bindDl.
\end{minted}

\section{Application 1: Quicksort}
\label{sec:quicksort}

In this section, we formalize the setting of the derivation of
quicksort by \cite{mu2020flops}.
This derivation relies crucially on nondeterminism: it uses the plus
monad (see Sect.~\ref{sec:plusMonad}), several reasoning steps amount
to make nondeterministic computations commute (see
Sect.~\ref{sec:commutativity}), and the derivation goal is expressed
in terms of refinement which is itself defined using nondeterminism.
In Sect.~\ref{sec:quicksort_derivation}, we state the proof goal of
the derivation of quicksort. We observe in particular that only
stating the goal requires care with dependent type to establish
termination.

\subsection{Plus monad and plus array monad}
\label{sec:plusMonad}

First, we extend the hierarchy of Sect.~\ref{sec:hier} with the
plus monad.  It extends the nondeterminism monad of Sect.~\ref{sec:nondeterminism}.

\begin{table}
\centering
\caption{The third set of laws satisfies by the plus monad (see
  Sect.~\ref{sec:plusMonad}) (\coqin{[~]} is the notation for
  nondeterministic choice, Sect.~\ref{sec:nondeterminism})}
\label{table:plusMonad}
\begin{tabular}{|l|l|}
\hline
\coqin{left_zero}    & \coqin{forall A B (f : A -> M B), fail A >>= f = fail B} \\
\coqin{right_zero}   & \coqin{forall A B (m : M A), m >> fail B = fail B} \\
\coqin{left_distributivity} & \coqin{forall A B (m1 m2 : M A) (f : A -> M B),} \\
                 & \coqin{  m1 [~] m2 >>= f = (m1 >>= f) [~] (m2 >>= f)} \\
\coqin{right_distributivity} & \coqin{forall A B (m : M A) (f1 f2 : A -> M B),} \\
                 & \coqin{  m >>= (fun x => f1 x [~] f2 x) =} \\
                 & \coqin{  (m >>= f1) [~] (m >>= f2)} \\
\hline
\end{tabular}
\end{table}

We define the plus monad following \cite{pauwels2019mpc} and
\cite{mu2020flops} (see \citep[Sect.~2]{mu2020flops}).  It extends a
basic monad with two operators: failure and nondeterministic choice.
These are the same operators as the nondeterminism monad.
It however satisfies more laws than the nondeterminism monad: (1)
failure and choice form a monoid, (2) choice is idempotent and
commutative, and (3) failure and choice interact with \coqin{bind} according
to the laws of Table~\ref{table:plusMonad}.
We take advantage of existing monads to implement the plus
monad. Indeed, we observe that most laws are already available.  The monad
\coqin{failMonad} (Sect.~\ref{sec:nondeterminism}) provides the
\coqin{left_zero} law.
The monad \coqin{failR0Monad} is used to define of the interface of
backtrackable
state~\citep[Sect.~3.2]{affeldt2019mpc} and in
\citep{affeldt2020types}; it introduces the \coqin{right_zero} law.
The monad \coqin{altMonad} (Sect.~\ref{sec:nondeterminism}) introduces nondeterministic choice and the
\coqin{left_distributivity} law.
The monad \coqin{altCIMonad} extends \coqin{altMonad} with
commutativity and idempotence of nondeterministic choice.
In other words, only the right-distributivity law is missing from
previous work.

\begin{figure}
\centering
\begin{tikzpicture}
\begin{scope}[every node/.style={draw,fill=white,rounded corners=0.3em},
              every path/.style={-stealth}]

\node (functor)  {\coqin{Functor}} ;
\node (monad) [below of=functor] {\coqin{Monad}} ;
\node (fail) [below right of=monad,xshift=1em] {\coqin{MonadFail}} ;
\node (alt) [below left of=monad,xshift=-1em] {\coqin{MonadAlt}} ;
\node (nondet) [below right of=alt,xshift=1em] {\coqin{MonadNondet}} ;
\node (altCI) [left of=nondet,xshift=-1.5cm] {\coqin{MonadAltCI}} ;
\node (failR0) [right of=nondet,xshift=1.5cm] {\coqin{MonadFailR0}} ;
\node (nondetCI) [below of=nondet,xshift=-1.5cm] {\coqin{MonadCINondet}} ;
\node (preplus) [right of=nondetCI,xshift=1.5cm] {\coqin{MonadPrePlus}} ;
\node (plus) [below of=preplus,xshift=-1cm] {\coqin{MonadPlus}} ;

\draw (functor) -- (monad) ;
\draw (monad) -- (fail) ;
\draw (monad) -- (alt) ;
\draw (fail) -- (nondet) ;
\draw (fail) -- (failR0) ;
\draw (alt) -- (nondet) ;
\draw (alt) -- (altCI) ;
\draw (altCI) -- (nondetCI) ;
\draw (nondet) -- (nondetCI) ;
\draw (nondet) -- (preplus) ;
\draw (failR0) -- (preplus) ;
\draw (preplus) -- (plus) ;
\draw (nondetCI) -- (plus) ;
\end{scope}
\end{tikzpicture}
\caption{Hierarchy of effects for the plus monad}
\label{fig:hier_plus}
\end{figure}
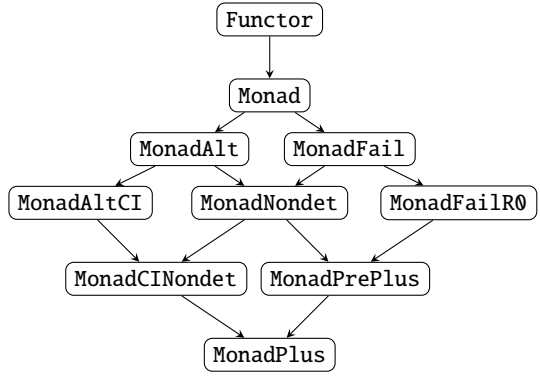

We therefore implement the plus monad by extending the monads
mentioned above with the
right-distributivity law as in Fig.~\ref{fig:hier_plus}. First, we define the intermediate
\coqin{prePlusMonad} by adding right-distributivity to the combination
of \coqin{nondetMonad} and \coqin{failR0Monad} (recall that
\coqin{alt} is the identifier behind the notation \coqin{[~]}).
\begin{minted}[bgcolor=mygray]{ssr}
HB.mixin Record isMonadPrePlus (M : UU0 -> UU0)
    of MonadNondet M & MonadFailR0 M :=
  { alt_bindDr : BindLaws.right_distributive (@bind M) alt }.

#[short(type=prePlusMonad)]
HB.structure Definition MonadPrePlus := {M of isMonadPrePlus M & }.  
\end{minted}

Then, \coqin{plusMonad} is defined as the combination of
\coqin{nondetCIMonad} and \coqin{prePlusMonad}:
\begin{minted}[bgcolor=mygray]{ssr}
#[short(type=plusMonad)]
HB.structure Definition MonadPlus := {M of MonadCINondet M & MonadPrePlus M}.  
\end{minted}

\paragraph*{Plus array monad}
The plus array monad is a straightforward
extension~\citep[Sect. 5]{mu2020flops}: it combines the plus monad of
this section with the array monad of Sect.~\ref{sec:arrayMonad}.
\begin{minted}[bgcolor=mygray]{ssr}
#[short(type=plusArrayMonad)]
HB.structure Definition MonadPlusArray (S : UU0) (I : eqType) :=
  { M of MonadPlus M & isMonadArray S I M}.
\end{minted}

\subsection{Commutativity of nondeterministic computations}
\label{sec:commutativity}

In monadic equational reasoning, it often happens that one needs to
show that two computations commute, in particular in presence of nondeterminism.
The following predicate \citep[Def 4.2]{mu2019tr3} defines the
commutativity of two computations \coqin{m} and \coqin{n} in \monae{}:
\begin{minted}[bgcolor=mygray,escapeinside=88]{ssr}
Definition commute {M : monad} A B
    (m : M A) (n : M B) C (f : A -> B -> M C) : Prop :=
  m >>= (fun x => n >>= (fun y => f x y)) =
  n >>= (fun y => m >>= (fun x => f x y)) :> M _.
\end{minted}
The need to show commutativity occurs in particular when dealing with
state (be it as in a state monad or in an array monad) and
nondeterminism. For example, commutation is possible when a
computation in the plus array monad is actually using only
nondeterministic operations and therefore does not access the array.
To capture computations that are actually using only nondeterministic
operations, we define a predicate that characterizes syntactically
nondeterminism monads. They are written with the following
(higher-order abstract \citep{pfenning1988pldi}) syntax where each
constructor reflects the eponymous monadic construct:
\begin{minted}[bgcolor=mygray,escapeinside=88]{ssr}
(* Module SyntaxNondet. *)
Inductive t : Type -> Type :=
| ret : forall A, A -> t A
| bind : forall B A, t B -> (B -> t A) -> t A
| fail : forall A, t A
| alt : forall A, t A -> t A -> t A.
\end{minted}
Let \coqin{sem} be a function that computes the semantics of the above
syntax in the nondeterminism monad:
\begin{minted}[bgcolor=mygray,escapeinside=88]{ssr}
Fixpoint sem {M : nondetMonad} {A} (m : t A) : M A :=
  match m with
  | ret A a => Ret a
  | bind A B m f => sem m >>= (sem \o f)
  | fail A => fail (* operator of the failure monad *)
  | alt A m1 m2 => sem m1 [~] sem m2
  end.
\end{minted}
Using the above syntax, we can for example write a predicate that
captures computations in a \coqin{plusMonad} that are semantically
just computations in the nondeterminism monad:
\begin{minted}[bgcolor=mygray,escapeinside=88]{ssr}
Definition plus_isNondet {M : plusMonad} A (n : M A) := {m | sem m = n}.
\end{minted}
Recall that the notation \coqin{{m | P m}} is for dependent pairs, as 
already explained in Sect.~\ref{sec:dassert}.


Eventually, it becomes possible to prove by induction on the syntax
that two computations \coqin{m} and \coqin{n} in the plus monad
commute when \coqin{m} only uses nondeterministic operators:
\begin{minted}[bgcolor=mygray,escapeinside=88]{ssr}
Context {M : plusMonad}.
Lemma plus_commute A (m : M A) B (n : M B) C (f : A -> B -> M C) :
  plus_isNondet m -> commute m n f.
\end{minted}
Using this lemma, to show that two computations in the plus monad
commute, it suffices to show that one of the two satisfies the
predicate \coqin{plus_isNondet}. When a computation does satisfy
this predicate, it is good practice to prove it right away and register
this fact as a hint in \coq{} so that the application of the \coqin{plus_lemma}
can be applied without even generating subgoals. Section~\ref{sec:stmt_quicksort}
provides a concrete example.

\subsection{Quicksort derivation}
\label{sec:quicksort_derivation}

This section formalizes the setting of \cite{mu2020flops}.  First, we
specify sorting functions. Second, we define quicksort using the array
monad.  Finally, we state the goal of the derivation of quicksort
using refinement.

\subsubsection{Nondeterministic computation of permutations}
\label{sec:qperm}

We formalize a function \coqin{qperm} that computes
nondeterministically permutations. In the following, \coqin{M} is
assumed to be of type \coqin{altMonad}
(Sect.~\ref{sec:nondeterminism}).

The first step is to define a function that splits a list
nondeterministically.  This is a structurally recursive function that
can be encoded using the \coqin{Fixpoint} command of \coq.  Recall that
\coqin{[~]} is the notation for nondeterministic choice.
\begin{minted}[bgcolor=mygray]{ssr}
Fixpoint splits s : M (seq A * seq A) :=
 if s isn't h :: t then
  Ret ([::], [::])
 else
  splits t >>= (fun xy => Ret (h :: xy.1, xy.2) [~] Ret (xy.1, h :: xy.2)).  
\end{minted}
The return type of the \coqin{splits} function is \coqin{M (seq A *
  seq A)}.  We now write another function \coqin{splits_bseq} with the
same semantics but whose return type contains size information that is
useful to establish the termination of functions calling it.
The return type of this new function  is \coqin{M
  ((size s).-bseq A * (size s).-bseq A)}, where \coqin{s} is the input
list and \coqin{n.-bseq A} is the type of lists of size less than or
equal to \coqin{n}\footnote{This type of \newterm{bounded-size lists} comes
from the \mathcomp{} library~\citep[file \coqin{tuple.v}]{mathcomp}.}.
\begin{minted}[bgcolor=mygray]{ssr}
Fixpoint splits_bseq s : M ((size s).-bseq A * (size s).-bseq A) :=
 if s isn't h :: t then
  Ret ([bseq of [::]], [bseq of [::]])
 else
  splits_bseq t >>= (fun '(x, y) =>
   Ret ([bseq of h :: x], widen_bseq (leqnSn _) y) [~]
   Ret (widen_bseq (leqnSn _) x, [bseq of h :: y])).
\end{minted}
Provided that one ignores the notations and lemmas about bounded-size
lists, the body of this definition is the same as the original
\coqin{splits}. The notation \coqin{[bseq of [::]]} is for an empty
list seen as a bounded-size list. The lemma \coqin{widen_bseq}
captures the fact that a \coqin{m.-bseq T} list can be seen as an
\coqin{n.-bseq T} list as long as \coqin{m <= n}:
\begin{minted}{ssr}
Lemma widen_bseq T m n : m <= n -> m.-bseq T -> n.-bseq T.
\end{minted}
Since \coqin{leqnSn n} is a proof of \coqin{n <= n.+1}, we understand
that \coqin{widen_bseq (leqnSn _)} turns a \coqin{n.-bseq} list into
an \coqin{n.+1.-bseq} list. The notation \coqin{[bseq of x :: ys]} is
a \mathcomp{} idiom that triggers type inference using canonical structures to
build a \coqin{n.+1.-bseq} list using the fact that \coqin{ys} is
itself a \coqin{n.-bseq} list.

We can now write a function \coqin{qperm} that computes permutations
nondeterministically.
For this purpose, we use the \coq{} \coqin{Equations} command
\citep{equations,sozeau2019icfp} that provides support to prove the
termination of functions whose recursion is not structural.
\begin{minted}[bgcolor=mygray]{ssr}
Equations? qperm (s : seq A) : M (seq A) by wf (size s) lt :=
| [::] => Ret [::]
| x :: xs => splits_bseq xs >>=
  (fun '(ys, zs) => liftM2 (fun a b => a ++ x :: b) (qperm ys) (qperm zs)).
\end{minted}
The annotation \coqin{by wf (size s) lt} indicates that the relation
between the sizes of lists is well-founded.
The function \coqin{liftM2} is a generic monadic function that lifts a
function \coqin{h} of type \coqin{A -> B -> C} to a monadic function
of type \coqin{M A -> M B -> M C}.
Once the \coqin{Equations} command is processed, \coq{} asks for a
proof that the size of the arguments are indeed decreasing.  This fact
is provable thanks to the additional type information returned by
\coqin{splits_bseq}.  Under the hood, \coq{} uses the accessibility
predicate \citep[Chapter 15]{bertot2004coq}.
Note that \coqin{qperm} is not the most obvious definition for the
task of generating nondeterministically permutations, but it is a good
fit to specify quicksort \citep[Sect. 3]{mu2020flops}.

\paragraph*{Specification of sorting functions}

The function \coqin{qperm} provides a way to specify sorting
functions. Indeed, it suffices to generate all the permutations
and filter the sorted ones.  Let use assume that \coqin{M} has
type \coqin{plusMonad} (Sect.~\ref{sec:plusMonad}) and that \coqin{T}
is an ordered type (as defined in \mathcomp).  We can use the generic
\mathcomp{} predicate \coqin{sorted} and define an obviously
correct sorting ``algorithm'' using \coqin{qperm}, the Kleisli symbol
\coqin{>=>}, and the assertion \coqin{assert} of Sect.~\ref{sec:dassert}:
\begin{minted}[bgcolor=mygray]{ssr}
Local Notation sorted := (sorted <=%O).

Definition slowsort : seq T -> M (seq T) := qperm >=> assert sorted.
\end{minted}

\subsubsection{In-place quicksort}

We can now formally define quicksort following \cite{mu2020flops}.

The partition step is performed by the function \coqin{ipartl}, which
uses the array monad (Sect.~\ref{sec:arrayMonad}):
\begin{minted}[bgcolor=mygray]{ssr}
Fixpoint ipartl p i ny nz nx : M (nat * nat) :=
  if nx is k.+1 then
    aget (i + (ny + nz)) >>= (fun x =>
      if x <= p then
        aswap (i + ny) (i + ny + nz) >> ipartl p i ny.+1 nz k
      else
        ipartl p i ny nz.+1 k)
  else Ret (ny, nz).
\end{minted}
The call \coqin{ipartl p i ny nz nx} partitions the subarray ranging
from index \coqin{i} (included) to index \coqin{i + ny + nz + nx} (excluded)
and returns the sizes of the two partitions. For the sake of
explanation, we can think of the contents of this subarray as a list
\coqin{ys ++ zs ++ xs}; \coqin{ys} and \coqin{zs} are the two
partitions and \coqin{xs} is yet to be partitioned; \coqin{ny} and
\coqin{nz} are the sizes of \coqin{ys} and \coqin{zs}.
At each iteration, the first element of \coqin{xs} (i.e., the element
at index \coqin{i + ny + nz}) is read and compared with the pivot
\coqin{p}. If it is smaller or equal, it is swapped with the element
following \coqin{ys} and partition proceeds with a \coqin{ys} enlarged
by one element (see Sect.~\ref{sec:arrayMonad} for the \coqin{aswap} function).
Otherwise, partition proceeds with a \coqin{zs} enlarged by one
element.

Quicksort is not structurally recursive and its definition therefore requires
care to convince \coq{} that it is really terminating.
Here, we formalize it using the \coqin{Fix} definition from the
\coqin{Coq.Init.Wf} module for well-founded fixpoint of the standard
library of \coq{}\footnote{Contrary to Sect.~\ref{sec:qperm}, we have
  not been able to use the \coqin{Equations} command here for
  technical reasons. The proof of termination of \coqin{iqsort} relies
  on type information carried by the return type of \coqin{dipartl}
  that the default setting of \coqin{Equations} fails to preserve.
  The \coqin{Program}/\coqin{Fix} is essentially the manual version of
  the \coqin{Equations} approach, it is less user-friendly but in practice it seems more widely applicable.}.
First, we write using the \coqin{Program} command for dependent type
programming \cite[Chapter Program]{coq} an intermediate function
\coqin{iqsort'} that implements the same logic as the intended
quicksort function except that it does not call itself recursively but
instead calls a parameter function (\coqin{f} at line \ref{line:f}
below).
The function \coqin{iqsort'} takes as input a pair of an
index and of a size (\coqin{ni}); it is a computation of the
\coqin{unit} type.
The parameter function \coqin{f} takes as an additional argument a
proof that the size of the input list is decreasing. These
corresponding proofs appear as holes (the \coqin{_} syntax) to be
filled by the user once the declaration is processed.
The code selects a pivot (line \ref{line:aget_pivot}), calls the
partition function (line~\ref{line:dipartl}), swaps two cells (line
\ref{line:aswap}), and then calls the parameter function on the
partitioned arrays:
\begin{minted}[bgcolor=mygray,numbers=left,xleftmargin=1.5em,escapeinside=88]{ssr}
Program Fixpoint iqsort' ni
    (f : forall mj, mj.2 < ni.2 -> M unit) : M unit := 8\label{line:f}8
  match ni.2 with
  | 0 => Ret tt
  | n.+1 => aget ni.1 >>= (fun p => 8\label{line:aget_pivot}8
            dipartl p ni.1.+1 0 0 n >>= (fun nynz => 8\label{line:dipartl}8
              let ny := nynz.1 in
              let nz := nynz.2 in
              aswap ni.1 (ni.1 + ny) >> 8\label{line:aswap}8
              f (ni.1, ny) _ >> f (ni.1 + ny.+1, nz) _))
  end.
\end{minted}
This formalization of in-place quicksort is actually a computation in
the plus array monad which is the only array monad that provides the
failure operator in the hierarchy of \monae{}.

The partition function is not exactly the \coqin{ipartl} function that
we explained at the beginning of this section but a
dependently-typed version \coqin{dipartl} that extends its return type
to a dependent pair of type \coqin{dipartlT}:
\begin{minted}[bgcolor=mygray]{ssr}
Definition dipartlT y z x :=
  {n : nat * nat | (n.1 <= x + y + z) && (n.2 <= x + y + z)}.  
\end{minted}
The parameters \coqin{y}, \coqin{z}, and \coqin{x} are the sizes of
the lists input to \coqin{ipartl} and this dependent type captures at the level of types that
the sizes returned by the partition function are smaller than the size
of the array being processed. The dependently-typed version of
\coqin{ipartl} is obtained by means of the predicate for
dependently-typed assertions of Sect.~\ref{sec:dassert}:
\begin{minted}[bgcolor=mygray]{ssr}
Definition dipartl p i y z x : M (dipartlT y z x) :=
  ipartl p i y z x >>=
    dassert [pred n | (n.1 <= x + y + z) && (n.2 <= x + y + z)].  
\end{minted}

Finally, the wanted actual \coqin{iqsort} function can be written
using \coqin{Fix}. This requires a (trivial) proof that the order
chosen for the measure is well-founded:
\begin{minted}[bgcolor=mygray,numbers=left,xleftmargin=1.5em,escapeinside=88]{ssr}
Lemma well_founded_lt2 :
  well_founded (fun x y : nat * nat => x.2 < y.2).
Proof. (* see 8\citep[file {\tt{}example\us{}iquicksort.v}]{monae}8 *) Qed.

Definition iqsort : nat * nat -> M unit :=
  Fix well_founded_lt2 (fun _ => M unit) iqsort'.
\end{minted}

The example of in-place quicksort is an example of the unfortunate but
unavoidable fact that programs must be terminating to be
shallow-embedded and verified in a proof assistant. Understanding the
support such as the \coqin{Equations} command or the \coqin{Fix}
construct in \coq{} as well as the need to sometimes reflect size
information at the level of types is important in practice and the
in-place quicksort provides a good example.
In theory, \newterm{sized types} have been shown to be useful to guarantee
the termination of programs such as quicksort~\citep{barthe2008csl}.
However, as of today, it appears that users of
the \coq{} proof assistant still need a support library to prove termination manually
since there are indications that sized types for \coq{} might not be practical~\citep{chan2023jfp}.

\subsubsection{Statement of the derivation of quicksort}
\label{sec:stmt_quicksort}

To state the correctness of \coqin{iqsort}, \cite{mu2020flops} define
a notion of program refinement that relates two monadic computations
using nondeterministic choice:
\begin{minted}[bgcolor=mygray]{ssr}
Definition refin {M : altMonad} A (m1 m2 : M A) : Prop := m1 [~] m2 = m2.
Notation "m1 `<=` m2" := (refin m1 m2).
\end{minted}
As the notation indicates, \coqin{m1 `<=` m2} represents a
relationship akin to set inclusion, which means that the result of
\coqin{m1} is included in the one of \coqin{m2}: we say that
\coqin{m1} \newterm{refines} \coqin{m2}.


The specification of the derivation of quicksort can now be written as
a refinement relation between, on the one hand, a program that writes
a list to memory (see Sect.~\ref{sec:arrayMonad}) and then calls
\coqin{iqsort}, and, on the other hand, a call to \coqin{slowsort} (see Sect.~\ref{sec:qperm})
followed by a program that writes the result to memory \citep[Eqn
12]{mu2020flops}:
\begin{minted}[bgcolor=mygray]{ssr}
Lemma iqsort_slowsort i xs :
  writeList i xs >> iqsort (M := M) (i, size xs) `<=`
  slowsort xs >>= writeList i.
\end{minted}

\def\ys{\textit{ys}}
\def\zs{\textit{zs}}
\def\zsp{\textit{zs'}}
\def\writeList{\textit{writeList}}
\def\cat{+\!+}
\def\swap{\textit{swap}}

The complete formal proof can be found in \citep[file
\coqin{example_iquicksort.v}]{monae}. Let us illustrate this formal proof with an
excerpt. In the course of deriving in-place quicksort,
\cite{mu2020flops} mutate an array $\ys\cat\zs\cat[\textbf{x}]$
to $\ys\cat[\textbf{x}]\cat\zsp$ where $\zsp$ is a permutation of $\zsp$
by swapping $\textbf{x}$ with the leftmost elements of $\zs$.
This is formalized by the following refinement relation \citep[Eqn 11, page 12]{mu2020flops}:
$$
\begin{array}{c}
\bindseq{\writeList\;(i+|\ys|)\;(\zs\cat [\textbf{x}])}{\swap}\;(i+|\ys|)\;(i+|\ys|+|\zs|) \\
\subseteq\\
\bindop{\textit{perm}\;\zs}{\lambda\zsp{}.\;\writeList\;(i+|\ys|)\;([\textbf{x}]\cat \zsp)}.
\end{array}
$$
In \monae{}, we formalize the above relation as the following lemma
(where \coqin{M} is the plus array monad):
\begin{minted}[bgcolor=mygray]{ssr}
Lemma refin_writeList_rcons_cat_aswap (i : nat) x (ys zs : seq E) :
  writeList i (rcons (ys ++ zs) x) >>
    aswap (M := M) (i + size ys) (i + size (ys ++ zs))
  `<=`
  qperm zs >>= (fun zs' => writeList i (ys ++ x :: zs')).
\end{minted}
There is almost a one-to-one match with the pencil-and-paper version
just above (the only difference is that we found it slightly more
practical to have the index \coqin{i} as a parameter instead of an
arithmetic expression).
Incidentally, the proof of the above lemma provides an application of
commutativity (Sect.~\ref{sec:commutativity}):

\noindent
\begin{tabular}{l}
\mintinline[fontsize=\footnotesize]{ssr}{qperm zs >>= (fun s => writeList i ys >> writeList (i + size ys) (x :: s))} \\
\inbra{\mintinline[fontsize=\footnotesize]{ssr}{rewrite (plus_commute (qperm zs))//.}} \\
\mintinline[fontsize=\footnotesize]{ssr}{writeList i ys >> (qperm zs >>= (fun s => writeList (i + size ys) (x :: s)))}
\end{tabular}

\noindent This proof step can be performed as a single tactic provided that \coqin{qperm} has
been proved to satisfy the \coqin{plus_isNondet} predicate of Sect.~\ref{sec:commutativity}
and that this fact has been registered as a hint in \coq{}.


% ...






\section{Application 2: ML programs with references}
\label{sec:typed_store}

In order to verify
\ocaml{} programs, we extend \monae{} with a new monad and its equational
theory so that programs generated by
\coqgen{}~\citep{garrigue2022types}, an \ocaml{} backend that outputs
\coq{} code, can be verified using monadic equational reasoning.
We call this monad the \newterm{typed-store monad\/} because it consists
essentially of a mutable typed store.
The original goal of \coqgen{} was ensuring the soundness of \ocaml{}
type inference by exploiting the richness of \coq's type system.
Using this approach, we can effectively reuse the output of \coqgen{},
instead of discarding it as a mere witness of type checking,
and harness it as a target for formal verification.
In this section, we do not give a full account of the typed-store
monad, as its theory is still in flux, but explain our methodology to
develop a new monad, and demonstrate how it can handle a non trivial
example involving cyclic lists.

\subsection{Representation of \ocaml{} types}

The typed store having heterogeneous contents, we need a concrete
representation of \ocaml{} types in order to handle the dynamic typing
required to access it. In \coqgen, this is achieved through an
inductive type \coqin{ml_type} of syntactic type representations (as
plain terms), and a computable function \coqin{coq_type}, interpreting
them into \coq{} types in the style of a Tarski
universe~\citep{martinlof1984}, which can be abstracted as the
following module interface.
\begin{minted}[bgcolor=mygray]{ssr}
Parameter ml_type : Set.
Variant loc : ml_type -> Type := mkloc T : nat -> loc T.
Parameter coq_type : forall N : Type -> Type, ml_type -> Type.
\end{minted}
The \coqin{loc} identifier above is for memory locations.
The interpretation function
\coqin{coq_type}~\citep{garrigue2022types}
is parameterized by a monad \coqin{N}, used to translate function types.
Concrete definitions for \coqin{ml_type} and \coqin{coq_type} are
generated by the \coqgen{} compiler.
 The presence of the Tarski universe
is the main difference with other transpilers from functional
programming languages to \coq{}, such as {\tt
  coq-of-ocaml}~\citep{claret2018phd} and {\tt
  hs-to-coq}~\citep{spector2018icfp}.

In \coqgen, the concrete monad used to run programs is obtained as a
fixpoint involving \coqin{coq_type}, which requires bypassing the
so-called strict-positivity requirement of inductive types. From the
point of view of type-soundness, this is a reasonable choice. However,
once we intend to use this framework for proof, it seems preferable to
avoid working in an inconsistent setting, so that, in \monae, we will
restrict ourselves to programs that do not store functions that
themselves access the store.

\subsection{Designing an equational theory from an implementation}

As \coqgen{} itself is built around a monad which models OCaml
computations, it seemed natural to use \monae{} for reasoning with it.
While \coqgen{} supports a large subset of \ocaml{}, including
references and exceptions, here we limit ourselves to references,
with reference creation, access, and update.

\subsubsection{Interface for effects}
In \monae, these effects are introduced as follows, starting with our
Tarski universe:
\begin{minted}[bgcolor=mygray]{ssr}
HB.mixin Structure isML_universe (ml_type : Type) := {
  eqclass : Equality.class_of ml_type ;
  coq_type : forall N : Type -> Type, ml_type -> Type ;
  ml_nonempty : ml_type ;
  val_nonempty : forall N, coq_type N ml_nonempty }.

#[short(type=ML_universe)]
HB.structure Definition ML_UNIVERSE := {ml_type & isML_universe ml_type}.

HB.mixin Record isMonadTypedStore (M : Type -> Type) (N : monad) of Monad M
  := {
  cnew : forall {T}, coq_type N T -> M (loc T) ;
  cget : forall {T}, loc T -> M (coq_type N T) ;
  cput : forall {T}, loc T -> coq_type N T -> M unit ;
  ... }.
\end{minted}
Note that, in order to preserve the soundness of our logic, we do not
take a fixpoint, so that this interface uses two different monads:
\coqin{M} for computations using the typed store, and \coqin{N} for
computations by values in the store~\footnote{Currently the relation between
  these two monads is not yet clear.  It seems that
  the existence of a monad morphism $\coqin{N} \to \coqin{M}$ is desired,
  but at the time of this writing, we have not investigated it.}

\subsubsection{Definition of the model}
\label{sec:typed_store_model}

The next step is to provide a model for this monad.
We started with the model from \coqgen{}, but soon realized that it
would be difficult to use it to prove laws, as it required too many
internal invariants.
Fortunately, we could quickly come up with an alternative model, whose
semantics is essentially identical, but which does not require explicit
invariants.
\begin{minted}[bgcolor=mygray]{ssr}
Variables (MLU : ML_universe) (N : monad).
Local Notation coq_type := (@coq_type MLU). (* the coq_type field of MLU *)
Local Notation ml_type := (MLU : Type).     (* the ml_type field of MLU *)

Record binding :=
  mkbind { bind_type : ml_type; bind_val : coq_type N bind_type }.

Definition M : UU0 -> UU0 :=
  MS (seq binding) [the monad of option_monad].

Let cnew T (v : coq_type N T) : M (loc T) :=
  fun st => let n := size st in
            Ret (mkloc T n, rcons st (mkbind (v : coq_type N T))).

Let cget T (r : loc T) : M (coq_type N T) :=
  fun st =>
    if nth_error st (loc_id r) is Some (mkbind T' v) then
      if coerce T v is Some u then Ret (u, st) else fail
    else fail.
\end{minted}
Here we just reuse the state monad-transformer of \monae{} (Sect.~\ref{sec:stateT}), applied to
the option monad, to allow computations to fail.
Interestingly, failure is only an internal feature of the model, and
does not appear in the interface: correctly translated OCaml programs
never fail on memory access, and none of the laws we present here
mentions failure\footnote{If we add an effect \coqin{crun} that
discards state, and returns only the result of a computation, failure
becomes visible in the interface. However, OCaml does not provide such
a function.}.
The state itself is
a list of dynamically typed values (\coqin{binding}) combining a type
representation and a value of this type. The type \coqin{loc} will
contain a position in this list.
The definitions of \coqin{cnew} and \coqin{cget} are introduced by
\coqin{Let} as we only need them to define an instance of our interface.
In the implementation of \coqin{cnew}, we generate a new reference
cell by adding a new value at the end of the list, using the \coqin{rcons}
operation. Since we use the length of the list as location, we can be
sure that this location was
invalid before calling \coqin{cnew}. We also show the code for
\coqin{cget}, which demonstrates how we access the store using a
standard library function, \coqin{nth_error}, which returns
\coqin{None} if the index is out of the list. We then use 
\coqin{coerce T v}, which uses the decidable equality of
\coqin{ml_type} to coerce a value to an expected type, but again
returns \coqin{None} if the dynamic type representation for \coqin{v}
is not \coqin{T}. We can see that
\coqin{cget T r} will only succeed on a store \coqin{st} if \coqin{r} is
a valid location which contains a value of type \coqin{T}.
The implementation of \coqin{cput} is similar to \coqin{cget}.

\subsubsection{Extending the interface with laws}
We move to the next step of our methodology: from this model, we
infer useful laws which we will prove and add to the interface.

\paragraph*{Adapting an existing theory}
As a starting point, we mimicked the array monad, which already
contains laws for \coqin{aput} and \coqin{aget}. The only difference
is that now types may differ, so we only show rules
\coqin{cgetputskip}, \coqin{cputC} and \coqin{cputgetC}, as
\coqin{cputput}, \coqin{cputget}, \coqin{cgetget} and \coqin{cgetC}
are identical to their array monad counterparts.

\begin{table}
\centering
\caption{Laws adapted from the array monad}
\label{tbl:typed_array}
\begin{tabular}{|l|l|}
\hline
\coqin{cgetputskip} &
\(\bindop{\coqin{cget}~ r}{\coqin{cput}~ r} =
\bindseq{\coqin{cget}~ r}{\coqin{skip}}\) \\
\coqin{cputC} &
\(\bindseq{\coqin{cput}~r_1~s_1}{\coqin{cput}~r_2~s_2} =
\bindseq{\coqin{cput}~r_2~s_2}{\coqin{cput}~r_1~s_2}\) \\
& \hfill if \(\coqin{loc_id}~ r_1 \neq \coqin{loc_id}~ r_2 \vee
\coqin{JMeq}~s_1~s_2\) \\
\coqin{cputgetC} &
\(\bindseq{\coqin{cput}~r_1~s}{(\bindop{\coqin{cget}~r_2}{k})} =
\bindop{\coqin{cget}~r_2}{\lambda
  v.\bindseq{\coqin{cput}~r_1~s}{k~v}}\) \hspace{5ex} \\
& \hfill if \(\coqin{loc_id}~ r_1 \neq \coqin{loc_id}~ r_2\) \\
\coqin{cgetputC} &
\(\bindseq{\coqin{cget}~r_1}{\coqin{cput}~r_2~s} =
\bindseq{\coqin{cput}~r_2~s}{(\bindseq{\coqin{cget}~r_1}{\coqin{skip}})} \)
\\ \hline
\end{tabular}
\end{table}
The most interesting change may be \coqin{cgetputskip}. Since
our judgments do not include a context part, we cannot guarantee that
access to \coqin{r} will be valid for any store.
So the left-hand side might fail, and is not equivalent to \coqin{skip}.
As a first approximation, we replace it with a call to \coqin{cget},
which would exhibit the same failure, and ignore its result by
composing with \coqin{skip}.
We will see later that this can be seen as a
separate operation, which checks the validity of a reference.
For \coqin{cputC}, there are two changes. The first one is that we
have a specific notion of identity for locations, which allows us to
compare locations of different types. The second one is that, if we
want to check the equality of the updated values, we need to use the
John Major equality~\citep{mcbride2004jfp}, which is defined as false whenever the types
differ. The law \coqin{cputgetC} is almost unchanged, but we added a variant
of it, \coqin{cgetputC}, where the result of \coqin{cget} is ignored,
and which requires no side condition.

\begin{table}
\centering
\caption{Laws for \coqin{cnew} and \coqin{cchk} (excerpt)}
\label{tbl:cnew_cchk}
\begin{tabular}{|l|l|}
\hline
\coqin{cnewget} &
\(\bindop{\coqin{cnew}~s}{\lambda r. \bindop{\coqin{cget}~r}{k~r}} =
\bindop{\coqin{cnew}~s}{\lambda r. k~r~s}\) \\
\coqin{cnewput} &
\(\bindop{\coqin{cnew}~s}{\lambda r. \bindseq{\coqin{cput}~r~t}{k~r}} =
\bindop{\coqin{cnew}~t}{k}\) \\
\coqin{cnewchk} &
\(\bindop{\coqin{cnew}~s}{\lambda r. \bindseq{\coqin{cchk}~r}{k~r}} =
\bindop{\coqin{cnew}~s}{k}\) \\
\coqin{cchknewC} &
\(\bindseq{\coqin{cchk}~r_1}{(\bindop{\coqin{cnew}~s}{\lambda
    r_2. \bindseq{\coqin{cchk}~r_1}{k~r_2}})} =
\bindseq{\coqin{cchk}~r_1}{(\bindop{\coqin{cnew}~s}{k})} \) \\
\coqin{cchknewE} &
\(\bindseq{\coqin{cchk}~r_1}{(\bindop{\coqin{cnew}~s}{k_1})} =
\bindseq{\coqin{cchk}~r_1}{(\bindop{\coqin{cnew}~s}{k_2})}\)
\hspace{10ex} \\
& \hfill if \(k_1~r_2 = k_2~r_2\) for all $r_2$ s.t. \(\coqin{loc_id}~ r_1 \neq
\coqin{loc_id}~ r_2\) \\
\coqin{cchkputC} &
\(\bindseq{\coqin{cchk}~r_1}{\coqin{cput}~r_2~s} =
\bindseq{\coqin{cput}~r_2~s}{\coqin{cchk}~r_1} \) \\
\coqin{cgetputchk} &
\(\bindop{\coqin{cget}~ r}{\coqin{cput}~ r} =
\coqin{cchk}~ r\) \\
\hline
\end{tabular}
\end{table}

\paragraph*{Inferring new laws through experimentation}
We then went on and added rules for \coqin{cnew}. The intuition is
that they should be similar to rules for \coqin{cput}, as they define
the contents of a reference, but there appears a new difficulty, as
there is no way to refer to the location of this  reference before it
is created.
This is only after experimenting with sample programs that we realized
that this could be exploited the other way around: if a reference is
valid before the creation of another one, then they cannot be equal.
And since the only way to check validity is through an operation in
the monad, we introduced $\coqin{cchk}~ r$ as an abbreviation for
$\bindseq{\coqin{cget}~r}{\coqin{skip}}$.
We show in Table~\ref{tbl:cnew_cchk} a non-exhaustive list of laws
involving \coqin{cnew} and \coqin{cchk}. Note that some rules are
intended to be used in the opposite direction, for instance
\coqin{cnewchk} allows to introduce a new \coqin{cchk}, which
\coqin{cchknewC} duplicates under another \coqin{cnew}.
The law \coqin{cchknewE} uses the same mechanism to infer an inequation
$\coqin{loc_id}~r_1 \neq \coqin{loc_id}~r_2$ between locations, which
can in turn be assumed to prove that continuations $k_1$ and $k_2$ are
equal (possibly using rules \coqin{cputC} or \coqin{cputgetC}).

\subsection{Examples}
During this process, we verified a number of example programs
\cite[file \coqin{example_typed_store.v}]{monae}: cyclic lists,
fibonacci, and factorial, all implemented in \ocaml{} using
references, and transpiled using \coqgen.
Let us consider the verification in \coq{} of the following
implementation of cyclic lists in OCaml:
\begin{minted}{ocaml}
type 'a rlist = Nil | Cons of 'a * 'a rlist ref

let cycle a b =
  let r = ref Nil in
  r := Cons (a, ref (Cons (b, r)));
  r

let rtl r = match !r with Nil -> r | Cons (_, l) -> l
\end{minted}
Generating the corresponding \coq{} formalization is as easy as running
\coqin{ocamlc -c -coq cycle.ml}\footnote{For the implementation, see \citep{garrigue2022types}.}.
We only have to annotate uses of \coqin{coq_type} with the appropriate
monad (\coqin{M} or \coqin{N}, while \coqgen{} only uses \coqin{M}).
\begin{minted}[bgcolor=mygray]{ssr}
Inductive ml_type := ... | ml_rlist (_ : ml_type).
... (* Proof of decidable equality for ml_type *)

Inductive rlist (a : Type) (a_1 : ml_type) :=
  | Nil
  | Cons (_ : a) (_ : loc (ml_rlist a_1)).

Fixpoint coq_type (N : Type -> Type) (T : ml_type) : Type :=
  match T with
  | ml_bool => bool
  | ml_rlist T1 => rlist (coq_type N T1) T1
  | ml_ref T1 => loc T1
  | ...
  end.

(* Create a ML universe instance *)
HB.instance Definition _ := @isML_universe.Build ml_type
  (Equality.class ml_type_eqType) coq_type ml_bool (fun _ => false).

(* Use the corresponding interface *)
Variables (N : monad) (M : typedStoreMonad ml_type N).

Definition cycle T (a b : coq_type N T) : M (loc (ml_rlist T)) :=
  do r <- cnew (ml_rlist T) (Nil (coq_type N T) T);
  do l <-
  (do v <- cnew (ml_rlist T) (Cons (coq_type N T) T b r);
   Ret (Cons (coq_type N T) T a v));
  do _ <- cput r l; Ret r.

Definition rtl T (r : loc (ml_rlist T)) : M (loc (ml_rlist T)) :=
  do v <- cget r; match v with | Nil => Ret r | Cons _ l => Ret l end.
\end{minted}
Note that \coqin{rlist}, being defined before \coqin{coq_type}, has to
take two parameters corresponding to the same type, which are later
related by \coqin{coq_type}.

As for the correctness statement, we will only check that the list
created by \coqin{cycle} is indeed a cycle of length 2:
\begin{minted}[bgcolor=mygray]{ssr}
Lemma rtl_tl_self T (a b : coq_type N T) :
  do l <- cycle T a b; do l1 <- rtl l; rtl l1 = cycle T a b.
\end{minted}
We show the derivation of this equality in Table~\ref{tbl:rtl_tl_self},
where we use the \coqin{under} tactic~\citep{martindorel2019coq}.
Written as a proof script it is just 8 lines long.

We have also proved the same result for cycles of length $n$.
In that case, it requires several lemmas, which together take about 60
lines to prove. We hope to be able to eventually improve scalability.

\def\tactic#1{\text{\inbra{\coqin{#1}}}}
\begin{table}
\caption{Derivation of \coqin{rtl_tl_self}.
Freshly inserted subexpressions are underlined. \\
Horizontal lines denote a change of hypotheses: adding and removing
\coqin{r1r2}.
}
\label{tbl:rtl_tl_self}
\(\displaystyle
\begin{array}l
(\bindop{\coqin{cnew Nil}}{\lambda r.
  \bindop
      {(\bindop{\coqin{cnew}~(\coqin{Cons f}~r)}
        {\lambda v. \rettt~(\coqin{Cons t}~v))}}
      {\lambda l. \bindseq{\coqin{cput}~r~l}{\rettt~l}}}
{}) \\
\hfill
\bindop{}
{\lambda l. \bindop{\coqin{rtl}~ l}{\coqin{rtl}}} \\
\tactic{rewrite bindA -cnewchk. (* insert cchk *)} \\
\bindop{\coqin{cnew Nil}}{\lambda r.
  \bindseq{\underline{\coqin{cchk}~r}}
  {((\bindop
    {(\bindop{\coqin{cnew}~\dots}
      {\lambda v. \rettt~(\coqin{Cons t}~v)})}
    {\lambda l. \bindseq{\coqin{cput}~r~l}{\rettt~l}})}}
\\
\hfill
\bindop{}
{\lambda l. \bindop{\coqin{rtl}~ l}{\coqin{rtl}}}) \\
\tactic{under eq_bind => r1. (* go under binder *)} \\
\coqin{'Under[}~
  \bindseq{\coqin{cchk}~r}
  {((\bindop
    {(\bindop{\coqin{cnew}~\dots}
      {\lambda v. \rettt~(\coqin{Cons t}~v)})}
    {\lambda l. \bindseq{\coqin{cput}~r~l}{\rettt~l}}))}
\\
\hfill
\bindop{}
{\lambda l. \bindop{\coqin{rtl}~ l}{\coqin{rtl}}})~\coqin{]} \\
\tactic{rewrite bindA; under eq_bind do rewrite !bindA.} \\
\coqin{'Under[}~
  \bindseq{\coqin{cchk}~r}
    {(\bindop{\coqin{cnew}~\dots}
      {\lambda v.
        (\bindop{\rettt~(\coqin{Cons t}~v)}
        {\lambda l. \bindseq{\coqin{cput}~r~l}{\rettt~l}})}}
\\
\hfill
\bindop{}
{\lambda l. \bindop{\coqin{rtl}~ l}{\coqin{rtl}}})~\coqin{]} \\
\tactic{under cchknewE => r2 r1r2. (* deduce r1r2 from cchk >> cnew *)} \\
\xdotfill{2pt}[purple] context = \coqin{r1r2 : loc_id r1 != loc_id r2} \xdotfill{2pt}[purple] \\
\coqin{'Under[}~
\bindop{(\bindop{\rettt~(\coqin{Cons t}~r_2)}
  {\lambda l.\bindseq{\coqin{cput}~r_1~l}{\rettt~r_1}})}
{\lambda l. \bindop{\coqin{rtl}~ l}{\coqin{rtl}~l_1}}
~\coqin{]} \\
\tactic{rewrite bindretf bindA bindretf. (* substitutions *)} \\
\coqin{'Under[}~
\bindseq{\coqin{cput}~r_1~\underline{(\coqin{Cons t}~r_2)}}
  {(\bindop{\coqin{rtl}~\underline{r_1}}{\coqin{rtl}})}
~\coqin{]} \\
\tactic{rewrite bindA cputget. (* evaluate (rtl r1) *)}\\
\coqin{'Under[ }
\bindseq{\coqin{cput}~r_1 (\coqin{Cons t}~r_2)}
  {(\bindop{\underline{\rettt~r_2}}{\coqin{rtl}})} \coqin{ ]} \\
\tactic{rewrite bindretf. (* substitution *)} \\
\coqin{'Under[}~
\bindseq{\coqin{cput}~r_1~(\coqin{Cons t}~r_2)}
  {\underline{\coqin{rtl}~r_2}}
~\coqin{]} \\
\tactic{rewrite -bindA. (* decompose rtl *)} \\
\coqin{'Under[ }
\bindop{(\bindseq{\coqin{cput}~r_1~(\coqin{Cons t}~r_2)}{\underline{\coqin{cget}~r_2}})}
{\underline{\lambda v. \rettt~(\coqin{match}~v~\coqin{with ... end})}}
\coqin{ ]} \\
\tactic{rewrite cputgetC //. (* commute using r1r2 *)} \\
\coqin{'Under[ }
\bindop{\underline{\coqin{cget}~r_2}}{\lambda v.
  \bindseq{\coqin{cput}~r_1~(\coqin{Cons t}~r_2)}
         {\rettt~(\coqin{match}~v~\coqin{with ... end})}} \coqin{ ]}
\\
\tactic{over. (* leave cchknewE premise *)} \\ \xdotfill{2pt}[purple] context = empty \xdotfill{2pt}[purple] \\
\coqin{'Under[ }
\bindseq{\coqin{cchk}~r_1}
{(\bindop{\coqin{cnew}~(\coqin{Cons f}~r_1)}
{\lambda r_2.\coqin{cget}~r_2}} \\
\hfill
\bindop{}{\lambda v.
  \bindseq{\coqin{cput}~r_1~(\coqin{Cons t}~r_2)}
          {\rettt~(\coqin{match}~v~\coqin{with Nil} \Rightarrow r_2
            \mid \coqin{Cons _}~t \Rightarrow t~\coqin{end})}}) \coqin{ ]}
\\
\tactic{rewrite cnewget. (* substitute v in continutation *)} \\
\coqin{'Under[ }
\bindseq{\coqin{cchk}~r_1}
{(\bindop{\coqin{cnew}~(\coqin{Cons f}~r_1)}
{\lambda r_2.
  \bindseq{\coqin{cput}~r_1~(\coqin{Cons t}~r_2)}
          {\rettt~\underline{r_1}}})} \coqin{ ]}
\\
\tactic{over.} \\
\bindop{\coqin{cnew Nil}}{\lambda r_1.
\bindseq{\coqin{cchk}~r_1}
{(\bindop{\coqin{cnew}~(\coqin{Cons f}~r_1)}
{\lambda r_2.
  \bindseq{\coqin{cput}~r_1~(\coqin{Cons t}~r_2)}
          {\rettt~r_1}})}}
\\
\tactic{rewrite cnewchk.} \\
\bindop{\coqin{cnew Nil}}{\lambda r_1.
\bindop{\coqin{cnew}~(\coqin{Cons f}~r_1)}
{\lambda r_2.
  \bindseq{\coqin{cput}~r_1~(\coqin{Cons t}~r_2)}
          {\rettt~r_1}}}
\\
\end{array}
\)
\end{table}

\section{Related Work}
\label{sec:related_work}

\subsection{Formalization of monads in \coq{}}

Monads have been widely used to model programming languages with
effects in the \coq{} proof assistant.

The main motivation is the verification of programs.  For example,
monads have been used in \coq{} to verify low-level systems
\citep{jomaa2018scp}, to verify effectful Haskell programs
\citep{christiansen2019haskell}, or for the modular verification of
low-level systems based on free monads \citep{letan2018fm}.
More directly related to the application of Sect.~\ref{sec:quicksort},
Sakaguchi provides a formalization of the quicksort algorithm in
\coq{} using the array state monad
\citep[Sect.~6.2]{sakaguchi2020scp}. His formalization is primarily
motivated by the generation of efficient executable code. This makes
for an intricate definition of quicksort (for example, all the
arguments corresponding to indices are bounded). Though his framework
does not prevent program verification
\citep[Sect.~4]{sakaguchi2020scp}, it seems difficult to reuse it for
monadic equational reasoning (the type of monads is specialized to
state/array and there is no hierarchy of monad interfaces).

Monads have also been used in \coq{} to reason
about the meta-theory of programming languages. For example,
\cite{delaware2013icfp} formalize several monads and monad
transformers in \coq{}, each one associated with a so-called feature
theorem.  When monads are combined, these feature theorems can then be
combined to prove type soundness

There are of course formalizations of monads in other proof
assistants. To pick one example that can be easily compared with a
formalization in \monae, one can find a formalization of the Monty
Hall problem in Isabelle/HOL~\citep{cock2012ssv} using the pGCL
programming language (to compare with
\citep[Sect.~2.3]{affeldt2021jfp}).

The idea to use packed classes~\citep{garillot2009tphols} to represent
a hierarchy of monad interfaces has been experimented
in~\citep{affeldt2019mpc}. Packed classes are known to scale up to
deep hierarchies of interfaces but their direct usage involves a mix
of techniques that lead to verbose code. The tool
\hb{}~\citep{cohen2020fscd} provides automation to hide technical
details from the user. It is also more robust. Indeed, we discovered
that our previous work~\citep[Fig. 1]{affeldt2020types} lacked an
intermediate interface, which required us to insert some type
constraints for type inference to succeed (see \citep{hbmissingjoin}
for details). \hb{} detects such omissions automatically.
The better-known type classes have been reported to replace packed
classes in many situations so they might provide an alternative to
formalize hierarchies of monad interfaces; they have been used for
this purpose to a lesser extent in related work (see for example the
accompanying material of \citep{mu2020flops}).

The application of Sect.~\ref{sec:quicksort} is a formalization of
\cite{mu2020flops} which comes with an Agda formalization as
accompanying material. The latter contains axiomatized facts
\citep[Table 1]{saito2022mpc}, including the termination of quicksort,
that are arguably orthogonal to the issue of quicksort derivation but
that reveals practical issues directly related to monadic equational
reasoning. We completed (and actually reworked from scratch) their
formalization using \monae{}.  To complete Mu and Chiang's
formalization, we needed in particular to formalize a thorough theory
of nondeterministic permutations \citep[Sect.~5.1]{saito2022mpc}. It
turns out that this is a recurring topic of monadic equational
reasoning. They are written in different ways depending on the target
specification: using nondeterministic selection
\citep[Sect.~4.4]{gibbons2011icfp}, using nondeterministic selection
and the function \coqin{unfoldM} \citep[Sect.~3.2]{mu2019tr3}, using
nondeterministic insertion \citep[Sect.~3]{mu2019tr2}, or using
\coqin{liftM2} \citep[Sect. 3]{mu2020flops}. The current version of
\monae{} has now a formalization of each.

\subsection{About monadic equational reasoning}

Gibbons and Hinze seem to be the first to synthesize monadic
equational reasoning as an approach
\citep{gibbons2011icfp,gibbons12utp,abousaleh2016}.  This viewpoint is
also adopted by other authors
\citep{oliveira2012jfp,chen2017netys,mu2019tr2,mu2019tr3,pauwels2019mpc,
  mu2020flops} that we have already mentioned in
Sect.~\ref{sec:introduction}.

Applicative functor is an alternative approach to represent effectful
computations. It has been formalized in Isabelle/HOL together with the
tree relabeling example \citep{lochbihler2016itp}. This work focuses
on the lifting of equations to allow for automation, while our
approach is rather the one of \mathcomp{}: the construction of a
hierarchy of mathematical structures backed up by a rich library of
definitions and lemmas to make the most out of the rewriting
facilities of \ssreflect.

\subsection{About monad transformers}

The idea to formalize monad transformers using packed classes was
experimented by~\cite{affeldt2020types} who formalize Jaskelioff's
theory of modular monad transformers.

Huffman formalizes three monad transformers in the Isabelle/HOL proof
assistant~\citep{huffman2012icfp}.  This experiment is part of a
larger effort to overcome the limitations of Isabelle/HOL type classes
to reason about Haskell programs that use (Haskell) type classes. In
the dependent type theory of \coq{}, we could formalize a much larger
theory, even relying on extra features of \coq{} such as
impredicativity and parametricity (see \citep{affeldt2020types} for
details).

Maillard proposes a meta language to define monad transformers in the
\coq{} proof assistant \citep[Chapter 4]{maillard2019phd}. It is an
instance implementation of one element of a larger framework to verify
programs with monadic effects using Dijskstra monads
\citep{maillard2019icfp}.  Their Dijkstra monads are based on
specification monads and are built using monad morphisms. Verification
of a monadic computation amounts to type it in \coq{} with the
appropriate Dijkstra monad.  Like Jaskelioff's theory of modular monad
transformers, the lifting of operations is one topic of this framework
but it does not go as far as the deep analysis of
Jaskelioff~\citep{jaskelioff2009esop,jaskelioff2009phd,jaskelioff2010}.

There are also formalizations of monads and their morphisms that focus
on the mathematical aspects, e.g., UniMath~\citep{UniMath}. However,
the link to the monad transformers of functional programming is not
done.

\subsection{About equational theories for ML with references}

The typed store monad, particularly when it is extended with a
\coqin{crun} operation to observe a result discarding the contents of
the store, is very similar to the ST monad of
Haskell~\citep{launchbury1994pldi}.
While we could not find a description of the equational theory of
the ST monad per se, there is a line of work on models for state using
the same operations.
This includes theories for local state, which allow partially or
completely discarding the state, and for global state, which do not
allow this. The state of the art for local state is
work by
\cite{Kammar2017lics}, which builds on work by \cite{staton2010fossacs} to
provide a theory for {\em full ground state}, where one can put
reference cells in the store, but not functions, i.e., basically the
same restriction as us, but we are currently limited to global state.
Note that they do not explicitly provide an equational theory, but
their model supports nominal equational reasoning in the style of
\cite{staton2010fossacs}, i.e., using judgments with a context that
makes valid references explicit. This is different from our plain
equational reasoning, which does not rely on such a context.
For global state, \cite{sterling2022denotational} recently proposed a
model for {\em higher-order state}, i.e., allowing functions
in the store, using synthetic guarded domain theory.
They went on to develop a program logic based on
it~\citep{aagaard2023mfps}, based on separation logic.
It allows equational reasoning to some extent, but again it relies on
a separation logic context when expressing equations. Moreover,
equations contain extra tick operations, to account for guardedness.

\section{Conclusions}
\label{sec:conclusion}

In this paper, we reported on an approach to formalize in the \coq{}
proof assistant an extensible hierarchy of effects with their
algebraic laws (Sect.~\ref{sec:hier}) as well as their models
(Sect.~\ref{sec:model_of_monads}).
The idea is to leverage on existing tools available for \coq{}: we use
\hb{} to formalize interfaces of effects and models of monads, and we
use \ssreflect{}'s rewriting tactic to reproduce proofs by monadic
equational reasoning.
We have illustrated the use of the framework in practice by
formalizing the setting of a derivation of in-place quicksort by
\cite{mu2020flops} in Sect.~\ref{sec:quicksort}.
We have also used our framework to design an original set of equations
so that \ocaml{} programs become amenable to monadic equational
reasoning in Sect.~\ref{sec:typed_store}.

Besides the two applications we have explained in this paper, we have
also reproduced a number of standard examples of monadic equational
reasoning:
\begin{itemize}
\item The tree relabeling example: This example originally motivated
  monadic equational reasoning \citep{gibbons2011icfp}. It amounts to
  show that the labels of a binary tree are distinct when the latter
  has been relabeled with fresh labels using the
  \coqin{freshMonad}. See \citep[file \coqin{example_relabeling.v}]{monae}.
\item The $n$-queens puzzle: This puzzle is used to illustrate the
  combination of state and nondeterminism. We have mechanized the
  relations between functional and stateful implementations
  \citep[Sections 6,7]{gibbons2011icfp}, as well as the derivation of
  a version of the algorithm using monadic hylo-fusion
  \citep[Sect.~5]{mu2019tr3}. See \citep[file
  \coqin{example_nqueens.v}]{monae}. Like the quicksort example, this
  example demonstrates the importance of commutativity lemmas
  (Sect.~\ref{sec:commutativity}).
\item The Monty Hall problem: We have mechanized the probability
  calculations for several variants of the Monty Hall problem
  \citep{gibbons2011icfp,gibbons12utp} using \coqin{probMonad}
  (Sect.~\ref{sec:probMonad}), \coqin{altProbMonad}
  (Sect.~\ref{sec:altprob_monad}), and \coqin{exceptProbMonad}
  \citep[file \coqin{example_monty.v}]{monae} \citep[Sect.~2.3]{affeldt2021jfp}.
\item Spark aggregation: Spark is a platform for distributed
  computing, in which the aggregation of data is therefore
  nondeterministic. Monadic equational reasoning can be used to sort
  out the conditions under which aggregation is actually deterministic
  \citep[Sect.~4.2]{mu2019tr2}. We have mechanized this result
  \citep[file \coqin{example_spark.v}]{monae}, which is part of a
  larger specification \citep{chen2017netys}.
\item The swap construction: This is an example of monad composition
  \citep{jones1993tr}.  Strictly speaking, this is not monadic
  equational reasoning: formalization does not require a mechanism
  such as packed classes. Yet, our framework proved adequate because
  it allows to mix in a single equation different units and joins
  without explicit mention of which monad they belong to \citep[file \coqin{monad_composition.v}]{monae}.
\end{itemize}

We believe that our approach is successful in the sense that it helps
find and fix errors in related work (as we have already explained in
Sect.~\ref{sec:introduction}), that proof scripts closely match their
paper-and-pencil counterparts (see, e.g., Fig.~\ref{fig:fastprod} and
Sect.~\ref{sec:stmt_quicksort}), and that it provides support to
investigate the design of new sets of equations (see
Sect.~\ref{sec:typed_store}). All these results have been aggregated
as one single \coq{} library \citep{monae} to provide reusable lemmas
and make monadic equational reasoning in the \coq{} proof assistant a
practical tool to verify effectful programs.

\subsection*{Acknowledgements}
The authors would like to thank Cyril Cohen and Enrico Tassi for their
assistance with \hb{}, Jeremy Gibbons, Ohad Kammar, and Shin-Cheng Mu
for their helpful input, and Yoshihiro Imai for his contribution
to~\cite{infotheo}. This work is partially based on previous work to
which David Nowak and Ayumu Saito have participated.



\bibliographystyle{JFPlike}
\bibliography{monae_journal.bib}

\begin{thebibliography}{62}

\bibitem[{Aagaard et~al.}(2023){Aagaard, Sterling, \&
  Birkedal}]{aagaard2023mfps}
{Aagaard, F.~L., Sterling, J. \& Birkedal, L.} (2023) {A denotationally-based
  program logic for higher-order store}. \textit{{Electronic Notes in
  Theoretical Informatics and Computer Science}}. {\bf 3}. Proceedings of the
  39th Conference on the Mathematical Foundations of Programming Semantics
  (MFPS XXXIX), Bloomington, IL, USA, June 19--23, 2023.

\bibitem[{Abou-Saleh et~al.}(2016){Abou-Saleh, Cheung, \&
  Gibbons}]{abousaleh2016}
{Abou-Saleh, F., Cheung, K.-H. \& Gibbons, J.} (2016) Reasoning about
  probability and nondeterminism. POPL workshop on Probabilistic Programming
  Semantics.

\bibitem[{Affeldt et~al.}(2020){Affeldt, Cohen, Kerjean, Mahboubi, Rouhling, \&
  Sakaguchi}]{affeldt2020ijcar}
{Affeldt, R., Cohen, C., Kerjean, M., Mahboubi, A., Rouhling, D. \& Sakaguchi,
  K.} (2020) Competing inheritance paths in dependent type theory: a case study
  in functional analysis. In 10th International Joint Conference on Automated
  Reasoning (IJCAR 2020), Paris, France, June 29--July 6. Springer. pp. 3--20.

\bibitem[{Affeldt et~al.}(2021){Affeldt, Garrigue, Nowak, \&
  Saikawa}]{affeldt2021jfp}
{Affeldt, R., Garrigue, J., Nowak, D. \& Saikawa, T.} (2021) A trustful monad
  for axiomatic reasoning with probability and nondeterminism. \textit{J.
  Funct. Program.} {\bf 31}(E17).

\bibitem[{Affeldt et~al.}(2023){Affeldt, Garrigue, \& Saikawa}]{saikawa2023coq}
{Affeldt, R., Garrigue, J. \& Saikawa, T.} (2023) Environment-friendly monadic
  equational reasoning for {OCaml}. The Coq Workshop 2023, Białystok, Poland,
  July 31, 2023. Abstract available at
  \url{https://coq-workshop.gitlab.io/2023/abstracts/coq2023_monadic-reasoning.pdf}.

\bibitem[{Affeldt and Nowak}(2020){Affeldt \& Nowak}]{affeldt2020types}
{Affeldt, R. \& Nowak, D.} (2020) Extending equational monadic reasoning with
  monad transformers. 26th International Conference on Types for Proofs and
  Programs ({TYPES} 2020), March 2--5, 2020, University of Turin, Italy.
  Schloss Dagstuhl - Leibniz-Zentrum f{\"{u}}r Informatik. pp. 2:1--2:21.

\bibitem[{Affeldt et~al.}(2019){Affeldt, Nowak, \& Saikawa}]{affeldt2019mpc}
{Affeldt, R., Nowak, D. \& Saikawa, T.} (2019) A hierarchy of monadic effects
  for program verification using equational reasoning. 13th International
  Conference on Mathematics of Program Construction ({MPC 2019}), Porto,
  Portugal, October 7--9, 2019. Springer. pp. 226--254.

\bibitem[{Baez and Shulman}(2010){Baez \& Shulman}]{baez2010}
{Baez, J.~C. \& Shulman, M.} (2010) Lectures on n-categories and cohomology.
  Towards Higher Categories. New York, NY. Springer New York. pp. 1--68.

\bibitem[{Barthe et~al.}(2008){Barthe, Gr{\'{e}}goire, \& Riba}]{barthe2008csl}
{Barthe, G., Gr{\'{e}}goire, B. \& Riba, C.} (2008) Type-based termination with
  sized products. 22nd International Workshop on Computer Science Logic ({CSL}
  2008), Bertinoro, Italy, September 16--19, 2008. Springer. pp. 493--507.

\bibitem[{Benton et~al.}(2000){Benton, Hughes, \& Moggi}]{benton2000appsem}
{Benton, N., Hughes, J. \& Moggi, E.} (2000) Monads and effects. Applied
  Semantics, International Summer School ({APPSEM} 2000), Caminha, Portugal,
  September 9-15, 2000, Advanced Lectures. Springer. pp. 42--122.

\bibitem[{Bertot and Cast{\'{e}}ran}(2004){Bertot \&
  Cast{\'{e}}ran}]{bertot2004coq}
{Bertot, Y. \& Cast{\'{e}}ran, P.} (2004) {\em Interactive Theorem Proving and
  Program Development---Coq'Art: The Calculus of Inductive Constructions\/}.
  Texts in Theoretical Computer Science. An {EATCS} Series. Springer.

\bibitem[{Chan et~al.}(2023){Chan, Li, \& Bowman}]{chan2023jfp}
{Chan, J., Li, Y. \& Bowman, W.~J.} (2023) Is sized typing for coq practical?
  \textit{J. Funct. Program.} {\bf 33}, e1.

\bibitem[{Chen et~al.}(2017){Chen, Hong, Leng{\'{a}}l, Mu, Sinha, \&
  Wang}]{chen2017netys}
{Chen, Y., Hong, C., Leng{\'{a}}l, O., Mu, S., Sinha, N. \& Wang, B.} (2017) An
  executable sequential specification for spark aggregation. 5th International
  Conference on Networked Systems ({NETYS} 2017), Marrakech, Morocco, May
  17--19, 2017. pp. 421--438.

\bibitem[{Cheung}(2017)]{cheungPhD2017}
{Cheung, K.-H.} (2017) {\em Distributive Interaction of Algebraic Effects\/}.
  Ph.D. thesis. Merton College, University of Oxford.

\bibitem[{Christiansen et~al.}(2019){Christiansen, Dylus, \&
  Bunkenburg}]{christiansen2019haskell}
{Christiansen, J., Dylus, S. \& Bunkenburg, N.} (2019) Verifying effectful
  {Haskell} programs in {Coq}. 12th {ACM} {SIGPLAN} International Symposium on
  Haskell (Haskell 2019), Berlin, Germany, August 18-23, 2019. {ACM}. pp.
  125--138.

\bibitem[{Claret}(2018)]{claret2018phd}
{Claret, G.} (2018) {\em Program in {Coq}. (Programmer en Coq)\/}. Ph.D.
  thesis. Paris Diderot University, France.

\bibitem[{Cock}(2012)]{cock2012ssv}
{Cock, D.~A.} (2012) Verifying probabilistic correctness in {Isabelle} with
  {pGCL}. 7th Conference on Systems Software Verification ({SSV} 2012), Sydney,
  Australia, 28--30 November 2012. pp. 167--178.

\bibitem[{Cohen et~al.}(2020){Cohen, Sakaguchi, \& Tassi}]{cohen2020fscd}
{Cohen, C., Sakaguchi, K. \& Tassi, E.} (2020) Hierarchy builder: Algebraic
  hierarchies made easy in {Coq} with {Elpi} (system description). 5th
  International Conference on Formal Structures for Computation and Deduction
  ({FSCD} 2020), June 29--July 6, 2020, Paris, France (Virtual Conference).
  Schloss Dagstuhl - Leibniz-Zentrum f{\"{u}}r Informatik. pp. 34:1--34:21.

\bibitem[{d.~S.~Oliveira et~al.}(2012){d.~S.~Oliveira, Schrijvers, \&
  Cook}]{oliveira2012jfp}
{d.~S.~Oliveira, B.~C., Schrijvers, T. \& Cook, W.~R.} (2012) {MRI:} modular
  reasoning about interference in incremental programming. \textit{J. Funct.
  Program.} {\bf 22}(6), 797--852.

\bibitem[{Delaware et~al.}(2013){Delaware, Keuchel, Schrijvers, \&
  d.~S.~Oliveira}]{delaware2013icfp}
{Delaware, B., Keuchel, S., Schrijvers, T. \& d.~S.~Oliveira, B.~C.} (2013)
  Modular monadic meta-theory. {ACM} {SIGPLAN} Int.\ Conf.\ on Functional
  Programming (ICFP 2013), Boston, MA, {USA}, September 25--27, 2013. pp.
  319--330.

\bibitem[{Garillot et~al.}(2009){Garillot, Gonthier, Mahboubi, \&
  Rideau}]{garillot2009tphols}
{Garillot, F., Gonthier, G., Mahboubi, A. \& Rideau, L.} (2009) Packaging
  mathematical structures. 22nd International Conference on Theorem Proving in
  Higher Order Logics (TPHOLs 2009), Munich, Germany, August 17--20, 2009.
  Springer. pp. 327--342.

\bibitem[{Garrigue and Saikawa}(2022){Garrigue \& Saikawa}]{garrigue2022types}
{Garrigue, J. \& Saikawa, T.} (2022) Validating {OCaml} soundness by
  translation into {Coq}. 28th International Conference on Types for Proofs and
  Programs (TYPES 2022), Nantes, France, 20--25 June, 2022. Abstract available
  at \url{https://www.math.nagoya-u.ac.jp/~garrigue/papers/types2022.pdf}.
  Implementation available as PR \url{https://github.com/COCTI/ocaml/pull/3}.

\bibitem[{Gibbons}(2012)]{gibbons12utp}
{Gibbons, J.} (2012) Unifying theories of programming with monads. Revised
  Selected Papers of the 4th International Symposium on Unifying Theories of
  Programming (UTP 2012), Paris, France, August 27--28, 2012. Springer. pp.
  23--67.

\bibitem[{Gibbons and Hinze}(2011){Gibbons \& Hinze}]{gibbons2011icfp}
{Gibbons, J. \& Hinze, R.} (2011) Just do it: simple monadic equational
  reasoning. 16th {ACM} {SIGPLAN} international conference on Functional
  Programming ({ICFP} 2011), Tokyo, Japan, September 19--21, 2011. {ACM}. pp.
  2--14.

\bibitem[{Gonthier and Mahboubi}(2010){Gonthier \& Mahboubi}]{gonthier2010jfr}
{Gonthier, G. \& Mahboubi, A.} (2010) An introduction to small scale reflection
  in {Coq}. \textit{J. Formaliz. Reasoning}. {\bf 3}(2), 95--152.

\bibitem[{{Hierarchy Builder}}(2021)]{hbmissingjoin}
{{Hierarchy Builder}}. (2021) Hierarchy builder wiki---missingjoin. Available
  at \url{https://github.com/math-comp/hierarchy-builder/wiki/MissingJoin}.

\bibitem[{Huffman}(2012)]{huffman2012icfp}
{Huffman, B.} (2012) Formal verification of monad transformers. {ACM} {SIGPLAN}
  International Conference on Functional Programming (ICFP 2012), Copenhagen,
  Denmark, September 9--15, 2012. {ACM}. pp. 15--16.

\bibitem[{{Infotheo}}(2018)]{infotheo}
{{Infotheo}}. (2018) {Infotheo}: A {Coq} formalization of information theory
  and linear error-correcting codes.
  \url{https://github.com/affeldt-aist/infotheo}. Authors: Reynald Affeldt,
  Manabu Hagiwara, Jonas S\'enizergues, Jacques Garrigue, Kazuhiko Sakaguchi,
  Taku Asai, Takafumi Saikawa, and Naruomi Obata. Last stable release: 0.6
  (2023).

\bibitem[{Jacobs}(2010)]{jacobs2010tcs}
{Jacobs, B.} (2010) Convexity, duality and effects. {IFIP} {TCS}. Springer. pp.
  1--19.

\bibitem[{Jaskelioff}(2009)]{jaskelioff2009esop}
{Jaskelioff, M.} (2009) Modular monad transformers. Programming Languages and
  Systems, 18th European Symposium on Programming ({ESOP} 2009), York, UK,
  March 22--29, 2009. Springer. pp. 64--79.

\bibitem[{Jaskelioff}(2009a)]{jaskelioff2009phd}
{Jaskelioff, M.} (2009a) {\em Lifting of Operations in Modular Monadic
  Semantics\/}. Ph.D. thesis. University of Nottingham. Available at
  \url{https://www.fceia.unr.edu.ar/~mauro/pubs/Thesis.pdf}.

\bibitem[{Jaskelioff and Moggi}(2010){Jaskelioff \& Moggi}]{jaskelioff2010}
{Jaskelioff, M. \& Moggi, E.} (2010) Monad transformers as monoid transformers.
  \textit{Theor. Comput. Sci.} {\bf 411}(51-52), 4441--4466.

\bibitem[{Jomaa et~al.}(2018){Jomaa, Nowak, Grimaud, \& Hym}]{jomaa2018scp}
{Jomaa, N., Nowak, D., Grimaud, G. \& Hym, S.} (2018) Formal proof of dynamic
  memory isolation based on {MMU}. \textit{Sci. Comput. Program.} {\bf 162},
  76--92.

\bibitem[{Jones and Duponcheel}(1993){Jones \& Duponcheel}]{jones1993tr}
{Jones, M.~P. \& Duponcheel, L.} (1993) Composing monads. Technical Report
  YALEU/DCS/RR-1004. Yale University.

\bibitem[{Kammar et~al.}(2017){Kammar, Levy, Moss, \& Staton}]{Kammar2017lics}
{Kammar, O., Levy, P.~B., Moss, S.~K. \& Staton, S.} (2017) A monad for full
  ground reference cells. 32nd Annual ACM/IEEE Symposium on Logic in Computer
  Science (LICS 2017), Reykjav\'{\i}k, Iceland. IEEE Press.

\bibitem[{Launchbury and Jones}(1994){Launchbury \& Jones}]{launchbury1994pldi}
{Launchbury, J. \& Jones, S. L.~P.} (1994) Lazy functional state threads. the
  {ACM} SIGPLAN'94 Conference on Programming Language Design and Implementation
  (PLDI), Orlando, Florida, USA, June 20--24, 1994. {ACM}. pp. 24--35.

\bibitem[{Letan et~al.}(2018){Letan, R{\'{e}}gis{-}Gianas, Chifflier, \&
  Hiet}]{letan2018fm}
{Letan, T., R{\'{e}}gis{-}Gianas, Y., Chifflier, P. \& Hiet, G.} (2018) Modular
  verification of programs with effects and effect handlers in coq. 22nd
  International Symposium on Formal Methods ({FM} 2018), Oxford, UK, July
  15--17, 2018. Springer. pp. 338--354.

\bibitem[{Liang et~al.}(1995){Liang, Hudak, \& Jones}]{liang1995popl}
{Liang, S., Hudak, P. \& Jones, M.~P.} (1995) Monad transformers and modular
  interpreters. 22nd {ACM} {SIGPLAN-SIGACT} Symposium on Principles of
  Programming Languages (POPL 1995), San Francisco, California, USA, January
  23--25, 1995. {ACM} Press. pp. 333--343.

\bibitem[{Lochbihler and Schneider}(2016){Lochbihler \&
  Schneider}]{lochbihler2016itp}
{Lochbihler, A. \& Schneider, J.} (2016) Equational reasoning with applicative
  functors. 7th International Conference on Interactive Theorem Proving (ITP
  2016), Nancy, France, August 22--25, 2016. Springer. pp. 252--273.

\bibitem[{Mahboubi and Tassi}(2022){Mahboubi \& Tassi}]{mathcompbook}
{Mahboubi, A. \& Tassi, E.} (2022) {\em Mathematical Components\/}. Zenodo.
  Available at \url{https://doi.org/10.5281/zenodo.3999478}.

\bibitem[{Maillard}(2019)]{maillard2019phd}
{Maillard, K.} (2019) {\em Principes de la Vérification de Programmes à
  Effets Monadiques Arbitraires\/}. Ph.D. thesis. Université PSL.

\bibitem[{Maillard et~al.}(2019){Maillard, Ahman, Atkey, Mart{\'{\i}}nez,
  Hritcu, Rivas, \& Tanter}]{maillard2019icfp}
{Maillard, K., Ahman, D., Atkey, R., Mart{\'{\i}}nez, G., Hritcu, C., Rivas, E.
  \& Tanter, {\'{E}}.} (2019) Dijkstra monads for all. \textit{{PACMPL}}. {\bf
  3}({ICFP}), 104:1--104:29.

\bibitem[{Martin-Dorel and Tassi}(2019){Martin-Dorel \&
  Tassi}]{martindorel2019coq}
{Martin-Dorel, E. \& Tassi, E.} (2019) {SSReflect} in {Coq} 8.10. The Coq
  Workshop 2019, Portland State University, Portland, OR, USA, September 8,
  2019. Presentation slides available at
  \url{https://staff.aist.go.jp/reynald.affeldt/coq2019/coqws2019-martindorel-tassi-slides.pdf}.

\bibitem[{Martin{-}L{\"{o}}f}(1984)]{martinlof1984}
{Martin{-}L{\"{o}}f, P.} (1984) {\em Intuitionistic type theory\/}. vol.~1 of
  {\em Studies in proof theory\/}. Bibliopolis.

\bibitem[{{MathComp}}(2023)]{mathcomp}
{{MathComp}}. (2023) The mathematical components repository. Available at
  \url{https://github.com/math-comp/math-comp}. Version 1.18.0.

\bibitem[{McBride and McKinna}(2004){McBride \& McKinna}]{mcbride2004jfp}
{McBride, C. \& McKinna, J.} (2004) The view from the left. \textit{J. Funct.
  Program.} {\bf 14}(1), 69--111.

\bibitem[{{Monae}}(2023)]{monae}
{{Monae}}. (2023) {Monae:} monadic effects and equational reasonig in {Coq}.
  \url{https://github.com/affeldt-aist/monae}. Authors: Reynald Affeldt, David
  Nowak, Takafumi Saikawa, Jacques Garrigue, Ayumu Saito, Celestine Sauvage,
  and Kazunari Tanaka. Last stable release: 0.5 (2023). This paper is about the
  master branch (to be released as version 0.6).

\bibitem[{Mu}(2019)]{mu2019tr2}
{Mu, S.} (2019) Equational reasoning for non-determinism monad: A case study of
  {Spark} aggregation. Technical Report TR-IIS-19-002. Academia Sinica.

\bibitem[{Mu}(2019a)]{mu2019tr3}
{Mu, S.} (2019a) Calculating a backtracking algorithm: An exercise in monadic
  program derivation. Technical Report TR-IIS-19-003. Academia Sinica.

\bibitem[{Mu and Chiang}(2020){Mu \& Chiang}]{mu2020flops}
{Mu, S. \& Chiang, T.} (2020) Declarative pearl: Deriving monadic quicksort.
  15th International Symposium on Functional and Logic Programming ({FLOPS}
  2020), Akita, Japan, September 14--16, 2020. Springer. pp. 124--138.

\bibitem[{Pauwels et~al.}(2019){Pauwels, Schrijvers, \& Mu}]{pauwels2019mpc}
{Pauwels, K., Schrijvers, T. \& Mu, S.} (2019) Handling local state with global
  state. 13th International Conference on Mathematics of Program Construction
  ({MPC} 2019), Porto, Portugal, October 7--9, 2019. Springer. pp. 18--44.

\bibitem[{Pfenning and Elliott}(1988){Pfenning \& Elliott}]{pfenning1988pldi}
{Pfenning, F. \& Elliott, C.} (1988) Higher-order abstract syntax. the {ACM}
  SIGPLAN'88 Conference on Programming Language Design and Implementation (PLDI
  1988), Atlanta, Georgia, USA, June 22--24, 1988. {ACM}. pp. 199--208.

\bibitem[{Saito and Affeldt}(2022){Saito \& Affeldt}]{saito2022mpc}
{Saito, A. \& Affeldt, R.} (2022) Towards a practical library for monadic
  equational reasoning in {Coq}. 14th International Conference on Mathematics
  of Program Construction ({MPC 2022}), Tbilisi, Georgia, September 26--28,
  2022. Springer. pp. 151--177.

\bibitem[{Sakaguchi}(2020)]{sakaguchi2020scp}
{Sakaguchi, K.} (2020) Program extraction for mutable arrays. \textit{Sci.
  Comput. Program.} {\bf 191}, 102372.

\bibitem[{Shan}(2018)]{shan2018tutorial}
{Shan, C.-C.} (2018) Equational reasoning for probabilistic programming. POPL
  2018 TutorialFest.

\bibitem[{Sozeau}(2009)]{equations}
{Sozeau, M.} (2009) Equations---a function definitions plugin. Available at
  \url{https://mattam82.github.io/Coq-Equations/}. Last stable release: 1.3
  (2022).

\bibitem[{Sozeau and Mangin}(2019){Sozeau \& Mangin}]{sozeau2019icfp}
{Sozeau, M. \& Mangin, C.} (2019) Equations reloaded: high-level
  dependently-typed functional programming and proving in coq. \textit{Proc.
  {ACM} Program. Lang.} {\bf 3}({ICFP}), 86:1--86:29.

\bibitem[{Spector{-}Zabusky et~al.}(2018){Spector{-}Zabusky, Breitner,
  Rizkallah, \& Weirich}]{spector2018icfp}
{Spector{-}Zabusky, A., Breitner, J., Rizkallah, C. \& Weirich, S.} (2018)
  Total {Haskell} is reasonable {Coq}. 7th {ACM} {SIGPLAN} International
  Conference on Certified Programs and Proofs ({CPP} 2018), Los Angeles, CA,
  USA, January 8--9, 2018. {ACM}. pp. 14--27.

\bibitem[{Staton}(2010)]{staton2010fossacs}
{Staton, S.} (2010) Completeness for algebraic theories of local state.
  Foundations of Software Science and Computational Structures. Berlin,
  Heidelberg. Springer Berlin Heidelberg. pp. 48--63.

\bibitem[{Sterling et~al.}(2022){Sterling, Gratzer, \&
  Birkedal}]{sterling2022denotational}
{Sterling, J., Gratzer, D. \& Birkedal, L.} (2022) Denotational semantics of
  general store and polymorphism. \textit{CoRR}. {\bf abs/2210.02169}.

\bibitem[{{The Coq Development Team}}(2023)]{coq}
{{The Coq Development Team}}. (2023) {\em The {Coq} Proof Assistant Reference
  Manual\/}. Inria. Available at \url{https://coq.inria.fr}. Version 8.18.0.

\bibitem[{Voevodsky et~al.}(2014){Voevodsky, Ahrens, Grayson, {\em
  et~al.\/}}]{UniMath}
{Voevodsky, V., Ahrens, B., Grayson, D. {\em et~al.\/}}. (2014) {UniMath---a
  computer-checked library of univalent mathematics}. Available at
  \url{https://github.com/UniMath/UniMath}.

\end{thebibliography}

\label{lastpage01}

\end{document}